\documentclass[english,twocolumn,amsmath,amssymb,superscriptaddress,showpacs]{revtex4-1}
\usepackage{mathptmx}
\usepackage[T1]{fontenc}
\usepackage[latin9]{inputenc}
\usepackage{amsmath}
\usepackage{amssymb}
\usepackage{graphicx}
\usepackage{esint}

\makeatletter

\providecommand{\tabularnewline}{\\}

\makeatother

\usepackage{babel}
\begin{document}

\title{Theoretical Study of Plasmonic Lasing in Junctions with many Molecules}

\author{Yuan Zhang }

\email{yzhang@phys.au.dk}

\author{Klaus M{\o}lmer}

\email{moelmer@phys.au.dk}

\address{Department of Physics and Astronomy, Aarhus University, Ny Munkegade
120, DK-8000 Aarhus C, Denmark}

\author{Volkhard May}

\email{may@physik.hu-berlin.de}

\address{Institute f{\"u}r Physik, Humboldt-Universit{\"u}t zu Berlin, Netwonstra{\ss}e
15, D-12489 Berlin, Germany}
\begin{abstract}
We calculate the quantum state of the plasmon field excited by an ensemble of molecular emitters, which are driven by exchange of electrons with metallic nano-particle electrodes. Assuming identical emitters that are coupled collectively to the plasmon mode but are otherwise subject to independent relaxation channels, we show that symmetry constraints on the total system density matrix imply a drastic reduction in the numerical complexity. For $N_{\text{m}}$ three-level molecules we may thus represent the density matrix by a number of terms scaling as $(N_{\rm m}+8)!/(8!N_{\rm m}!)$ instead of $9^{N_{\text{m}}}$, and this allows exact simulations of up to $N_{\text{m}}=10$ molecules. 
Our simulations demonstrate that many emitters compensate strong plasmon damping and lead to the population of high plasmon number states and a narrowed linewidth of the plasmon field. For large $N_{\text{m}}$, our exact results are reproduced by an approximate approach based on the plasmon reduced density matrix. With this approach, we have extended the simulations to more than $50$ molecules and shown that the plasmon number state population follows a Poisson-like distribution. An alternative approach based on nonlinear rate equations for the molecular state populations and the mean plasmon number also reproduce the main lasing characteristics of the system.
\end{abstract}


\maketitle

\section{Introduction}

Hybrid systems of metal nano-particles (MNP) and molecular
quantum emitters have attracted increased attention in
the last two decades \cite{YYin,PBerini}. In particular, nonlinear effects, related to the quantum nature
of (surface) plasmons, i.e. collective oscillation of conductance
band electrons in the MNP offer interesting lasing effects
in the so-called plasmonic nano-laser \cite{MANoginov,XGMeng,YJLu,WZhou,AYang,CZhang,RMMa,CYWu,YJLu1,QZhang,KDing,MTHill}.
Rather than the radiation mode in a normal laser cavity, the plasmonic nano-laser utilizes a
confined plasmon oscillation. Since the size of the nano-laser can be much smaller than the wavelength of the light generated by the plasmon,
the optical diffraction limit restricting the size of conventional
laser does not apply for the plasmonic nano-laser \cite{stockman}.

The rapid damping of plasmon excitations impedes the achievement of plasmonic lasing
\cite{GKeues} unless many quantum emitters concertedly transfer their energy to the MNP.
These quantum emitters should, in turn be excited, e.g., by optical pumping
\cite{MANoginov,XGMeng,YJLu,WZhou,AYang,CZhang,RMMa,CYWu,YJLu1,QZhang} or by electrical injection \cite{KDing,MTHill}.
In Ref.\cite{YZhang}, we proposed several approaches
to describe the optically pumped plasmonic nano-laser, noting that the analyses can be
readily extended to the electrically pumped nano-laser \cite{YZhang-1}.

To compensate the plasmon damping and to realize strong plasmon excitation,
we can either increase the excitation strength or the number of quantum emitters. The increased
strength raises the excited state probability of the emitters, which enables them to transfer
energy more efficiently to the MNP. However, it  may be hard to realize in experiments with a single emitter. If we turn to many quantum emitters, their mutual coupling can lead to the formation of exciton states,
which may have different transition energies compared to that of the isolated quantum emitters, i.e. inhomogeneous broadening  \cite{VNPustovit}. As a result, the energy transfer from the
quantum emitters to the MNP plasmon becomes inefficient, and the coupling to other plasmon modes may occur, which further complicates the situation \cite{GKeues}.

In this work we aim to prove the principle of plasmonic lasing with an idealized system, where many identical and non-interacting quantum emitters couple with one plasmon mode. This ideal system has been
studied with non-linear rate equations \cite{stockman,ASRosenthal,IEProtsenko,SWChang,SWChang2,VMParfenyev},
with a Fokker-Planck equation \cite{VMParfenyev2} as well as with a reduced density
matrix equation (RDM) \cite{MRichter,YZhang}. In most studies, the quantum emitters are treated as
two-level systems incoherently driven by an effective pump. Although this effective description can capture the main physics of the systems, it is unable to describe the real experiments.
For example, the optical pumping utilized in Ref. \cite{MANoginov,YJLu,XGMeng,RMMa} relies on two processes: a laser excitation of the quantum emitters to a higher excited state and a subsequent decay to a lower excited state. The electrical injection applied in Ref. \cite{KDing,MTHill} also requires the participation of intermediate electron states.

To go beyond the effective description, the theories developed so-far should be extended and the quantum emitters should be treated as multi-levels systems. Fortunately, the theory presented in \cite{YZhang} can be easily extended to three-level systems. Such an extension will be elaborated in the present article.
In particular, the numerical exact method to solve RDM equations, which was initially proposed by M. Richter and A. Knorr in \cite{MRichter} for many identical two-levels systems coupled to a cavity or plasmon mode, can be easily extended to the case of many identical multi-levels systems. This method actually provides another way to solve the dissipative Cavity-QED problem without using Dicke states \cite{UMartini} and therefore can be utilized to study many related problems, for example, sub-radiance  and super-radiance \cite{LMandel}, lasing \cite{SEHarris, MXu},  and collective behavior of spin-ensembles \cite {JHWesenberg, ZKurucz,BAChase}.

In the present article, we apply the extended theory to an electrically pumped molecular junction, cf. Fig.\ref{fig:scheme_molJunMNP}(a), and demonstrate that a laser with electrically pumped dye molecules is theoretially possible. Our study may also contribute to the exploration of plasmon lasers with an electrically pumped organic semiconductor layer, cf. Fig.\ref{fig:scheme_molJunMNP}(b). More processes are involved, for example interlayer electron-transfer, exciton formation etc., and the modeling of such a structure is beyond the scope of the present article. 
If realized, however, this kind of laser may replace the relative expensive semiconductor laser and may thus have strong impact on the industry \cite{FJDuarte, KHayashi}.

The article is organized as follows. In Sec. \ref{sec:model}, we introduce the model of the junction system.
In Sec. \ref{sec:general-approach}, we present
a general RDM approach to solve the system dynamics, the electric current signal and the optical emission spectrum. This approach was previously used to simulate junctions
with up to 5 molecules in \cite{YZhang-1}. In Sec. \ref{sec:approach-identical-molecules}, we develop
an approach for junctions with identical molecules, which all couple coherently to the plasmon mode but decay and decohere independently. The symmetry of the RDM is explored to drastically reduce the computational
effort, and simulations are carried out for junctions with up
to 10 molecules. In Sec. \ref{sec:approach-plasmon-RDM}, we eliminate the molecular degrees of freedom and introduce and solve an approximate master equation for the plasmon RDM, which readily applies for junctions with up to 50 molecules. In Sec. \ref{sec:approach-rate-equations}, we eliminate instead the plasmon state and obtain non-linear equations for the molecular RDM. This approach allows calculation of the mean plasmon number which we compare with the results obtained by the other methods. The paper ends with several concluding remarks and an outlook in Sec. \ref{sec:conclusions}.

\section{Molecular Junction Model \label{sec:model}}

\begin{figure}
\begin{centering}
\includegraphics[scale=0.95]{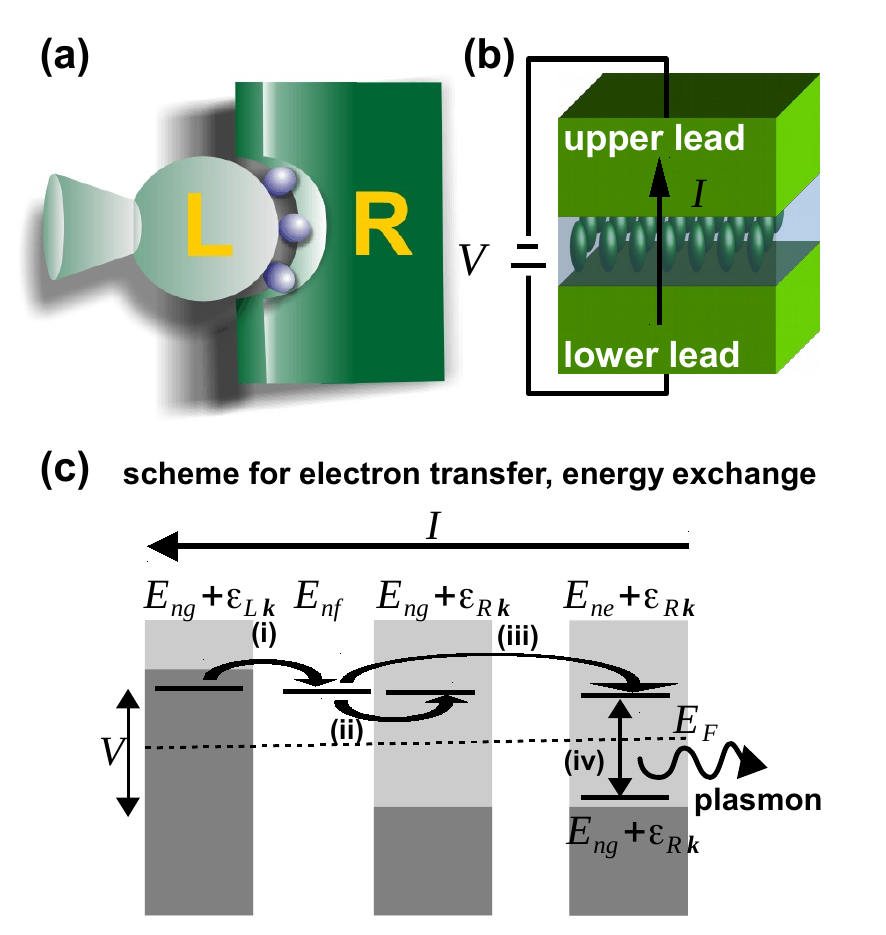}
\par\end{centering}

\caption{ \label{fig:scheme_molJunMNP} The physical system consists of molecules sandwiched between a spherical left lead supporting plasmon excitations and a right hollow lead, cf. panel (a). Panel (b) shows an alternative system with an organic semiconductor layer sandwiched between two metallic layers, supporting cavity plasmons between the layers. Schematic panel (c): light (dark) gray areas indicate unoccupied (occupied) electron states of the electrodes; boundaries of the areas are Fermi energies; the dashed line indicates the Fermi energies at zero-applied voltage. Energy conservation, cf. Eq. (\ref{eq:coupling-functions}), requires: for charging process (i),  energy of neutral molecule (ground state) plus that of electrons of the left lead $E_{ng} + \varepsilon_{L \bold k}$ is identical to  energy of charged molecule $E_{nf}$; for discharge processes (ii) and (iii), the energy $E_{nf}$ is equal to that of neutral molecule (excited or ground state) plus that of unoccupied electron states of the right lead $E_{ne} + \varepsilon_{R \bold k}$ or $E_{ng} + \varepsilon_{R \bold k}$ (shown separately for clarity); process (iv) shows energy exchange with the left lead plasmon. }
\end{figure}

The  molecular junction, formed by many molecules suspended between two metallic leads, is shown in Fig.\ref{fig:scheme_molJunMNP} (a). The left lead with a spherical form can support
plasmon excitations, and we assume that one of its plasmon modes is resonant with a molecular transition.
The right cavity-shaped lead may also support
plasmon excitations, which, we assume, are far off-resonance with the
left lead plasmon and the molecules. The molecules are placed
between the two leads and are well separated from each other so that
their excitonic coupling can be ignored. We assume that their transition dipole moments
are tangential to the surface of the left lead and that they couple resonantly with same strength to one
 single plasmon mode. Similar assumptions may be valid for layer configurations, cf.  Fig.\ref{fig:scheme_molJunMNP} (b).

The theoretical description of the junction follows Ref. \cite{YZhang-1}.  The molecular
Hamiltonian reads
\begin{equation}
H_{\text{mol}}=\sum_{n=1}^{N_{\text{m}}}\sum_{a_n}E_{na}\left|a_{n}\right\rangle \left\langle a_{n}\right|.\label{eq:mol_Ham}
\end{equation}
Here,  $N_{\text{m}}$ is the number
of molecules, and $a_{n}$ denotes the electronic state of the $n^{th}$ molecule. The set of relevant states includes the ground state $a_{n}=g_{n}$ and the excited state $a_{n}=e_{n}$
of the neutral molecule, as well as the  negatively charged
molecular (ground) state $a_{n}=f_{n}$, cf. Fig.\ref{fig:scheme_molJunMNP} (c). The transition dipole moment of the neutral molecules is $\mathbf{d}_{n}=d_{\text{mol}}\mathbf{e}_{n}$.

The dipole plasmons of the left spherical lead can be modeled as three degenerate
quantum harmonic oscillators \cite{GWeick} with the Hamiltonian
\begin{equation}
H_{\text{pl}}=\sum_{I}\hbar\omega_{\text{pl}}C_{I}^{+}C_{I}.\label{eq:plasmonHam}
\end{equation}
Here, $C_{I}^{+}$ and $C_{I}$ are creation and annihilation operator
of the plasmon excitations with excitation
energy $\hbar\omega_{\text{pl}}$ (the ground-state
energy is defined to be zero). We choose the plasmon modes such that the plasmon dipole moments read $\mathbf{d}_{I}=d_{\text{pl}}\mathbf{e}_{I}$, where
$\mathbf{e}_{I}$ are Cartesian unit vectors $I=x,y,z$. There is a further infinite set of multipole
plasmon oscillators, which however are not resonant with the molecules and are hence omitted from our analysis, cf. the discussion in Ref. \cite{YZhang,YZhang-1}.

The plasmons get excited by a resonant energy exchange with the molecular emitters, which in turn are excited through an electron transfer between the leads and the molecules \cite{VMay1,VMay2}, cf. Fig.\ref{fig:scheme_molJunMNP} (c). Note that this differs  from the non-resonant plasmon excitation through direct, inelastic electron transfer between metallic leads \cite{SWWu}. We assume that the electron transfer is weak and that the coupling of the molecular states to the electron continuum states in the leads can be treated by a master equation approach, see below.
The resonant energy exchange coupling is described by
\begin{equation}
H_{\text{mol-pl}}=\sum_{nI}\left(V_{nI}\left|e_{n}\right\rangle \left\langle g_{n}\right|C_{I}+V_{In}\left|g_{n}\right\rangle \left\langle e_{n}\right|C_{I}^{+}\right).\label{eq:molPl}
\end{equation}
Here, the coupling coefficient reads $V_{nI}=d_{\text{mol}}d_{\text{pl}}\kappa_{nI}/R_{n}^{3}$.
$R_{n}$ denotes the distance between the center of the molecule $n$
and that of the left lead, and $\mathbf{n}_{n}$ is the related unit
vector. The geometry factor takes the form $\kappa_{nI}=\left[\mathbf{e}_{n}\mathbf{e}_{I}\right]-3\left[\mathbf{e}_{n}\mathbf{n}_{n}\right]\left[\mathbf{n}_{n}\mathbf{e}_{I}\right]$. Although the coupling strength with higher plasmon modes may be larger than that with dipole plasmons, their corresponding energy transfer can be very weak if the molecules are off-resonant to them \cite{YZhang,YZhang-1}. This justifies the dipole-dipole interaction used here.

\section{General Approach Based on Reduced Density Matrix\label{sec:general-approach}}

In this section, we apply the open quantum system approach to investigate the
dynamics of the molecular junction.  The combined system of molecules and plasmon modes is described by the Hamiltonian $H_{\text{S}}=H_{\text{mol}}+H_{\text{pl}}+H_{\text{mol-pl}}$. The lead electron reservoirs influence the junction dynamics through incoherent processes between the neutral (ground and excited state) molecule and the ground negatively charged molecular state.  Considering the reservoir Hamiltonian $H_{\text{R}}$ and system-reservoir interaction $H_{\text{S-R}}$ specified in \cite{YZhang-3,YZhang-4}, we obtain a master equation for a density operator $\hat{\rho}$ (see below).  We describe the system with a density matrix $\rho_{\alpha \mu,\beta \nu} \equiv {\rm tr_S} \{ \hat{\rho}  \left|\beta\nu\right\rangle \left\langle \alpha\mu\right| \}$. The matrix is constructed in a complete basis formed by the product states
\begin{equation}
\left|\alpha\mu\right\rangle =\prod_{n} \left|a_{n}\right\rangle \prod_{I}\left|\mu_{I}\right\rangle ,\label{eq:BasisStates}
\end{equation}
The index $\alpha$ abbreviates the set of molecular states $\left\{ a_{1},...,a_{N_{\text{m}}}\right\} $, and the index $\mu$ abbreviates the set of Fock state quantum numbers $\left\{ \mu_{x},\mu_{y},\mu_{z}\right\} $. In our calculations, we include all the molecular states but truncate the plasmon multiple states at a maximum value.

\subsection{Equation of Motion for Reduced Density Matrix}

The equation of motion for $\hat{\rho}$ \cite{YZhang-1} reads
\begin{equation}
\frac{\partial}{\partial t}\hat{\rho}=-\frac{i}{\hbar}\left[H_{\text{S}},\hat{\rho}\right]_{-}-\mathcal{D}\left[\hat{\rho}\right],\label{eq:DensityOperator}
\end{equation}
where the system Hamiltonian $H_{\text{S}}$ was introduced
in the previous section and the dissipative superoperator $\mathcal{D}$
is chosen according to the following Lindblad-form
\begin{equation}
\mathcal{D} \left[ \hat{\rho} \right]=\frac{1}{2}\sum_{u}k_{u}\left(\left[\hat{L}_{u}^{+}\hat{L}_{u},\hat{\rho}\right]_{+}-2\hat{L}_{u}\hat{\rho}\hat{L}_{u}^{+}\right).\label{eq:dissipative}
\end{equation}
The plasmon damping with a total rate $\gamma_{\text{pl}}$ is included by identifying $\hat{L}_{u}$
as $C_{I}$ and $k_{u}$ as $\gamma_{\text{pl}}$. This rate is same for the three plasmon modes and includes both
the interaction with radiation field and with electron-hole pair excitations inside the left lead. If the plasmon modes of the right lead are also relevant (not the case for the configuration in Fig.\ref{fig:scheme_molJunMNP}(c)), we can also include their damping here.

Charging transitions into the charged molecular ground state may occur from both the ground and excited state of the neutral molecule. They are included in the treatment by choosing $\hat{L}_{u}=\left|f_{n}\right\rangle \left\langle b_{n}\right|$ and $k_{u}=k_{b\to f}^{\left(n\right)}$ with $b_{n}=g_{n},e_{n}$. The charging rates $k_{b\to f}^{\left(n\right)}$ take the form
\begin{eqnarray}
k_{b\to f}^{\left(n\right)} & =\sum_{X}k_{Xb\to f}^{\left(n\right)}= & \sum_{X}\Gamma_{Xbf}^{\left(n\right)}f_{\text{F}}\left(E_{fb}^{\left(n\right)}-\mu_{X}\right) \label{eq:charging_rate}
\end{eqnarray}
with the molecule-lead coupling function
\begin{equation}
\Gamma_{Xbf}^{\left(n\right)}=2\pi\sum_{\mathbf{k}s}\left|V_{X\mathbf{k}s}^{nbf}\right|^{2}\delta\left(\epsilon_{X\mathbf{k}}-E_{fb}^{\left(n\right)}\right)\label{eq:coupling-functions}
\end{equation}
multiplying the electron energy distribution $f_{\text{F}}$ in the lead $X$. In the above expression, $\mathbf{k}$ and $s$ denote the wave-vector and
spin of the electrons in the lead $X$. The coefficient $V_{X\mathbf{k}s}^{nbf}$
describes the amplitude of exchanging one electron between the lead $X=L,R$ and the
molecule $n$, accompanying a molecular transition between the neutral
state $\left|b_{n}\right\rangle $ and the singly charged state $\left|f_{n}\right\rangle $.
The molecular charging is possible if the energy $\epsilon_{X\mathbf{k}}$ of electrons
in the leads coincides with the charging
energy $E_{fa}^{\left(n\right)}=E_{nf}-E_{na}$, cf. Eq. (\ref{eq:coupling-functions}), which leads to the electron exchange scheme shown in Fig.\ref{fig:scheme_molJunMNP}(c).
The Fermi-distribution function $f_{\text{F}}$ in Eq. (\ref{eq:charging_rate})
ascertains that only the occupied electron states in the leads contribute to the molecular
charging. The lead chemical potentials assume the values $\mu_{X=L}=E_{\text{F}}+|e|V/2$
and $\mu_{X=R}=E_{\text{F}}-|e|V/2$ (for the case of a symmetrically
applied voltage). Here, $E_{\mathrm{F}}$ is the chemical potential
of the leads at zero-voltage bias, cf. the dashed line in Fig.\ref{fig:scheme_molJunMNP}(c).

Discharge of the molecules towards the leads is similarly included
by Lindblad terms with $\hat{L}_{u}=\left|b_{n}\right\rangle \left\langle f_{n}\right|$
and $k_{u}=k_{f\rightarrow b}^{\left(n\right)}$. The discharging rate
is obtained from Eq. (\ref{eq:charging_rate}) by replacing $f_{\text{F}}$
with $1-f_{\text{F}}$ (only the unoccupied electron states in the leads are available for
the molecular charge transfer).

Radiative decay of the excited molecular states, which is caused
by interaction with quantized radiation field,  can be also readily included in the equation (\ref{eq:dissipative}). Since
the molecular radiative decay rate is orders of magnitude
weaker than the molecular charging and discharge rates \cite{YZhang-1}, we are justified to ignore it in the present work.

\subsection{Current Formula}

The steady state current through the molecular junction is defined as the number
of electrons passing through the molecules per time. This current
can in turn be determined from the rate of the processes exchanging electrons between the molecules and one of the leads ($X$):
\begin{equation}
I_{X}=\sum_{n=1}^{N_{\text{m}}}\sum_{b=g,e}\left(k_{Xb\to f}^{\left(n\right)}P_{nb}-k_{Xf\to b}^{\left(n\right)}P_{nf}\right),\label{eq:current}
\end{equation}
where the charging and discharging rates, $k_{Xb\to f}^{\left(n\right)}$ and $k_{Xf\to b}^{\left(n\right)}$
were already introduced in the previous section, and $P_{ng}$, $P_{ne}$
and $P_{nf}$ are the populations of the neutral and the singly negatively charged molecular states $\left|g_{n}\right\rangle $,$\left|e_{n}\right\rangle $, and  $\left|f_{n}\right\rangle$. The molecular populations are directly obtained from the solution of the master equation:
\begin{align}
P_{nc} & \equiv\text{tr}_{\text{S}}\left\{ \hat{\rho}\left|c_{n}\right\rangle \left\langle c_{n}\right|\right\} =\sum_{\mu}\sum_{\alpha'}\rho_{\alpha'\mu,\alpha'\mu},
\end{align}
Here, the label $\alpha'=\left\{ a_{1},...,c_{n},...,a_{N_{\text{m}}}\right\} $
indicates the molecular product states, where the molecule $n$ is
in the electronic state $\left|c_{n}\right\rangle =\left|g_{n}\right\rangle$, $\left|e_{n}\right\rangle$ or $\left|f_{n}\right\rangle$.

\subsection{Emission Spectrum Formula}

The master equation also gives access to the steady state power spectrum $F\left(\omega\right)$ of the emitted radiation. This quantity is evaluated as the Fourier transform of the two-time correlation function of the emitting dipoles,
\begin{equation}
F\left(\omega\right)=\frac{4\omega^{3}}{3\pi c^{3}\hbar}\text{Re}\int_{0}^{\infty}dte^{-i\omega t}\sum_{A,B}\left[\mathbf{d}_{A}\mathbf{d}_{B}^{*}\right] \left < \hat{X}_{A}^{+}(t) \hat{X}_{B}(0) \right >.\label{eq:EmissionOperator}
\end{equation}
 The indices $A$ and $B$  indicate the contributions from the molecules and the plasmonic modes. Correspondingly, $\hat{X}_{A}^{+}$ and $\hat{X}_{B}$ are transition operators $\hat{X}_{A}=\left|\varphi_{ng}\right\rangle \left\langle \varphi_{ne}\right|$, or $\hat{X}_{A}=C_I$, while $\mathbf{d}_{A}$ and $\mathbf{d}_{B}^{*}$ denote the corresponding transition dipole moments.

According to the quantum regression theorem \cite{PMeystre}, the two-time correlation functions defined by the expectation values $\left < \hat{X}_{A}^{+}(t) \hat{X}_{B}(0) \right > \equiv \text{tr}_{\text{S}}\left\{ \hat{X}_{A}^{+}\hat{\sigma}\left(B;t\right)\right\}$ can be calculated by propagating the operator (matrix) $\hat{\sigma}\left(B;t\right)\equiv\mathcal{U}\left(t\right)\left[\hat{X}_{B}\hat{\rho}_{\text{ss}}\right]$
with the same time-evolution super-operator $\mathcal{U}\left(t\right)$ that propagates the density matrix according to Eq. (\ref{eq:DensityOperator}). The initial value for $\hat{\sigma}\left(B;t\right)$ is the product of the
steady-state reduced density matrix $\hat{\rho}_{\text{ss}}$ and the matrix expression for
$\hat{X}_{B}$,  
\begin{eqnarray}
 &  & \sigma_{\alpha\mu,\beta\nu}\left(B,0\right)=\text{tr}_{\text{S}}\left\{ \left|\beta\nu\right\rangle \left\langle \alpha\mu\right|\hat{X}_{B} \hat{\rho}_{\text{ss}}\right\} \nonumber \\
 & = & \sum_{\alpha'\mu'}\left\langle \alpha\mu\right|\hat{X}_{B}\left|\alpha'\mu'\right\rangle \rho_{\alpha'\mu',\beta\nu}^{\left(\text{ss}\right)}.
\end{eqnarray}
Here, $\rho_{\alpha'\mu',\beta\nu}^{\left(\text{ss}\right)}$
is the reduced density matrix of the junction at steady-state.

The master equation solutions for $\rho_{\alpha\mu,\beta\nu}$
and $\sigma_{\alpha\mu,\beta\nu}\left(B,t\right)$ have been obtained for junctions with up to $5$ molecules in \cite{YZhang-1}.
There, it was demonstrated that with increasing number of molecules, higher plasmon excited stat es get
populated and the emission becomes more narrowed. These results are similar
to the lasing operation observed in experiments \cite{MANoginov,YJLu,KDing}.
In fact, it is amplified spontaneous emission of plasmon rather than 
lasing since, on average, only about one quantum of plasmon is excited. To demonstrate
lasing, we should consider junctions with more molecules.
However, since the number $9^{N_{\text{m}}}\left(N_{\text{pl}}+1\right)^{2}$  of $\rho_{\alpha\mu,\beta\nu}$ exponentially increases with $N_{\text{m}}$ ($N_{\text{pl}}$ indicates the highest excited plasmon state
in the simulations), it is impossible to
carry out the desired simulations.  In the following sections we develop exact and approximate approaches that mitigate the nine-fold increase in computational
effort for each extra molecule included in the system.

\subsection{Parameters of Simulations}

Here we specify the system parameters used in our
simulations, see Table \ref{tab:para}. The left lead forms a spherical shape
with $20$ nm diameter, and the dipole plasmons have an excitation
energy of $\hbar\omega_{\text{pl}}=2.6$ eV, a transition dipole moment
of $d_{\text{pl}}=2925$ D, and a damping rate $\hbar\gamma_{\text{pl}}=57$
meV \cite{YZhang-3}. The molecules are positioned at the left lead surface
at a distance of $\Delta x_{\text{mol-MNP}}=2.5$ nm. The molecular
excitation energy shall be in complete resonance with the dipole plasmons,
i.e., $\hbar\omega_{n}=2.6$ eV. The molecular transition dipole moment
is chosen as $16$ D, which is in line with our previous study \cite{YZhang-1}.

\begin{table}
\caption{Physical parameters in our calculations (for further explanation see text)\label{tab:para}}
\centering{}%
\begin{tabular}{cc|cc}
\hline
$\hbar\omega_{n}$ & $2.6$ eV & $\Delta E_{10}$ & $1.3$ eV\tabularnewline
$d_{\text{mol}}$ & $16$ D & $\hbar\Gamma_{Ref}^{\left(n\right)}$ & $50$ meV\tabularnewline
$\hbar\omega_{\text{pl}}$ & $2.6$ eV & $\hbar\Gamma_{Lgf}^{\left(n\right)}$ & $30$ meV\tabularnewline
$\hbar\gamma_{\text{pl}}$ & $57$ meV & $\hbar\Gamma_{Lef}^{\left(n\right)},\hbar\Gamma_{Rgf}^{\left(n\right)}$ & $1$ meV\tabularnewline
$d_{\text{pl}}$ & $2925$ D & $V$ & $3$ V\tabularnewline
$\Delta x_{\text{mol-MNP}}$ & $2.5$ nm & $k_{B}T$ & $5$ meV\tabularnewline
\hline
\end{tabular}
\end{table}

To specify the molecular charging and discharging rate, we
introduce the so-called relative charged energy as $\Delta E_{10}=E_{nf}-E_{ng}-E_{\text{F}}$
(assumed to be identical for all the molecules, $E_{\text{F}}$ is Fermi-energy at zero applied voltage bias) \cite{VMay2}. Here, we set it as $\Delta E_{10}=\hbar\omega_{n}/2$. The applied voltage
is $3$ V, at which the excited state of the neutral molecule is populated
through electron transfer process \cite{YZhang-1}. The lead- and
state-dependent molecule-lead couplings are chosen as $\hbar\Gamma_{Ref}^{\left(n\right)}=50$
meV, $\hbar\Gamma_{Lgf}^{\left(n\right)}=30$ meV and $\hbar\Gamma_{Lef}^{\left(n\right)}=\hbar\Gamma_{Lgf}^{\left(n\right)}=1$ meV. As shown in \cite{YZhang-1}, these values are optimal for the
population inversion of the molecules and thus for the lead plasmon
excitation. The thermal energy entering in the Fermi-distribution
function is set as $k_{B}T=5$ meV (at low temperature).

\section{Approach for Junctions with Identical Molecules\label{sec:approach-identical-molecules}}

Next, a theoretical treatment of a junction with identical molecules
is developed. In this case, the symmetry reduces the number of independent elements in the 
density matrix $\rho_{\alpha\mu,\beta\nu}$ because many
matrix elements have identical values. The application of the symmetric effective parametrization 
to an ensemble of three-level system, as carried out here, generalizes earlier work  \cite{MRichter,YZhang} on the plasmonic nano-laser in which the emitters were modeled as two-level systems.

\begin{figure}
\begin{centering}
\includegraphics[scale=0.85]{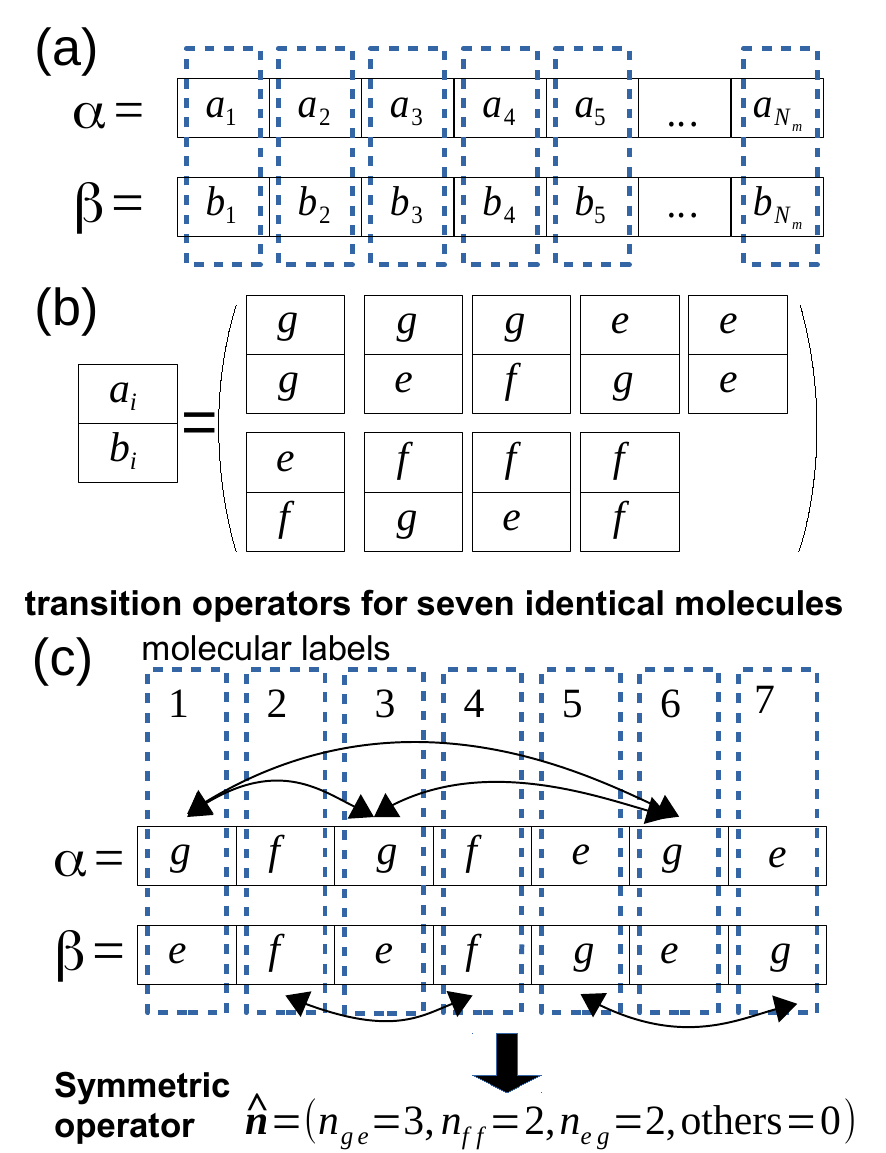}
\par\end{centering}

\caption{ \label{fig:symmetric-rdm} Transition operators $\left|\alpha\right\rangle \left \langle \beta \right|$ based on molecular product states $\left|\alpha\right\rangle ,\left|\beta\right\rangle $ are mapped to vectors $\mathbf{n}$. Panel (a): expansion of the molecular product states. Panel (b): possible combinations $(a_i,b_i)$ of quantum numbers related to individual molecules. Panel (c): some transition operators for seven identical molecules as one example; the group of molecules 1,3,6 is associated with the combination $(g,e)$, the group of molecules 2,4 with $(f,f)$, the group of molecules 5,7 with  $(e,g)$; the transition operators formed by exchanging the molecules in one group are identical; the set of those transition operators can by classified with the number of nine combinations in the panel (b) giving us a symmetric operator.\label{fig:mapping}}
\end{figure}

\subsection{Symmetric Reduced Density Matrix}

If all the molecules are identical, the original reduced density matrix (RDM) can
be represented by a symmetric RDM  according to the mapping $\rho_{\alpha\mu,\beta\nu} \equiv {\rm tr_S} \{\hat{\rho} \left|\alpha\right\rangle \left \langle \beta \right| \times \left| \mu \right\rangle \left \langle \nu \right| \}  \to  \rho_{ \mathbf{n}}^{\mu,\nu} \equiv {\rm tr_s} \{\hat{\rho}  \hat {\mathbf{n}} \times \left| \mu \right\rangle \left \langle \nu \right|\} $. Here,  the molecular transition operators $\left|\alpha\right\rangle \left \langle \beta \right|$ are mapped to symmetric operators $\hat {\mathbf{n}}$, cf. Fig.\ref{fig:mapping}. The value of the symmetric operators is defined as $\mathbf{n}=\left(n_{gg},n_{ge},n_{gf},n_{eg},n_{ee},n_{ef},n_{fg},n_{fe},n_{ff}\right)$ with nine positive integers in the range $\left[0,N_{\text{m}}\right]$, cf. Fig.\ref{fig:mapping} (b) .
In general, these integers can be written as $n_{cd}$ ($c,d=g,e,f$)
and can be determined by the following formula $n_{cd}=\sum_{l=1}^{N_{\text{m}}}\delta_{a_{l},c}\delta_{b_{l},d}$, cf. Fig.\ref{fig:mapping} (c) for systems with seven identical molecules.  It indicates the number of molecules, which are on the state $\left|c\right\rangle $
in the product state $\left|\alpha\right\rangle $ and simultaneously
on the state $\left|d\right\rangle $ in the product state $\left|\beta\right\rangle $.
Here, $a_{l}$ and $b_{l}$ are elements of the sets $\alpha$ and $\beta$, respectively.  The above definition naturally leads to $\sum_{c,d}n_{cd}=N_{\text{m}}.$ In fact, one symmetric RDM element represents a group of identical 
original RDM elements. Therefore, the treatment keeps all the information of systems without invoking any assumption.
 Because the above consideration is based on the product states rather than Dicke states, it can be readily applied for systems with multi-levels emitters \cite{YZhang-5}.

We shall refer to the number of elements  as the number of the vector $\mathbf{n}$.
We consider $N_{\text{m}}$ molecules as indistinguishable (identical)
balls and the nine components of $\mathbf{n}$ as nine distinguished
boxes. Then, the number of $\mathbf{n}$ is equal to the number of
possibilities to put these balls into the boxes. The latter is a well-known
combinatorial problem, and the result is $C_{n}^{k}\equiv n!/\left[k!\left(n-k\right)!\right]$
with $k+1$ boxes and $n+k$ balls. Therefore, the number of $\mathbf{n}$ is $C_{N_{\text{m}}+8}^{8}$.

For the special configuration shown in Fig.\ref{fig:scheme_molJunMNP},
only the plasmon mode $I=z$ interacts with the molecules. Therefore,
the indices $\mu,\nu$ in $\rho_{\alpha\mu,\beta\nu}$ and $\rho_{\mathbf{n}}^{\mu,\nu}$
are occupation numbers of the states $\left|\mu=\mu_{z}\right\rangle $
and $\left|\nu=\nu_{z}\right\rangle $. Due to the strong plasmon damping,
very high laying plasmon states are expected to be unpopulated. Therefore,
we can truncate the plasmon states in the simulations. We use $N_{\text{pl}}$
to indicate the highest plasmon excited state considered. Consequently,
the number of $\rho_{\mathbf{n}}^{\mu,\nu}$ is $n_{\text{tot}}=C_{N_{\text{m}}+8}^{8}\left(N_{\text{pl}}+1\right)^{2}$.
Notice that the number of $\rho_{\alpha\mu,\beta\nu}$ is $9^{N_{\text{m}}}\left(N_{\text{pl}}+1\right)^{2}$.
Obviously, the size of $\rho_{\mathbf{n}}^{\mu,\nu}$ is much smaller
than that of $\rho_{\alpha\mu,\beta\nu}$.

The equation of motion for $\rho_{\mathbf{n}}^{\mu,\nu}$ is obtained
by replacing $\rho_{\alpha\mu,\beta\nu}$ in its equation with the
corresponding $\rho_{\mathbf{n}}^{\mu,\nu}$ and noticing the definition
of $n_{cd}$. The final result is
\begin{eqnarray}
 &  & \frac{\partial}{\partial t}\rho_{\mathbf{n}}^{\mu,\nu}\nonumber \\
 & = & -i[\left(n_{eg}-n_{ge}\right)\omega_{eg}+\left(n_{fg}-n_{gf}\right)\omega_{fg}\nonumber \\
 & + & \left(n_{fe}-n_{ef}\right)\omega_{fe}]\rho_{\mathbf{n}}^{\mu,\nu}-i\omega_{\textrm{pl}}\left(\mu-\nu\right)\rho_{\mathbf{n}}^{\mu,\nu}\nonumber \\
 & + & i\nu_{\text{mol-pl}}\sum_{a=g,e,f}[\sqrt{\nu}n_{ag}\rho_{\left(n_{ag}-1,n_{ae}+1\right)}^{\mu,\nu-1}\nonumber \\
 & - & \sqrt{\mu+1}n_{ea}\rho_{\left(n_{ea}-1,n_{ga}+1\right)}^{\mu+1,\nu}+\sqrt{\nu+1}n_{ae}\rho_{\left(n_{ae}-1,n_{ag}+1\right)}^{\mu,\nu+1}\nonumber \\
 & - & \sqrt{\mu}n_{ga}\rho_{\left(n_{ga}-1,n_{ea}+1\right)}^{\mu-1,\nu}]\nonumber \\
 & - & \left(\gamma_{\textrm{pl}}/2\right)\left[\left(\mu+\nu\right)\rho_{\mathbf{n}}^{\mu,\nu}-2\sqrt{\left(\nu+1\right)\left(\mu\text{+1}\right)}\rho_{\mathbf{n}}^{\mu+1,\nu+1}\right]\nonumber \\
 & - & \sum_{b=g,e}\left(k_{b\to f}/2\right)[\sum_{a=g,e,f}\left(n_{ba}+n_{ab}\right)\rho_{\mathbf{n}}^{\mu,\nu}\nonumber \\
 & - & 2n_{ff}\rho_{\left(n_{ff}-1,n_{bb}+1\right)}^{\mu,\nu}]\nonumber \\
 & - & \sum_{b=g,e}\left(k_{f\to b}/2\right)[\sum_{a=g,e,f}\left(n_{fa}+n_{af}\right)\rho_{\mathbf{n}}^{\mu,\nu}\nonumber \\
 & - & 2n_{bb}\rho_{\left(n_{bb}-1,n_{ff}+1\right)}^{\mu,\nu}].\label{eq:molJunRhoN}
\end{eqnarray}
Here, $\hbar\omega_{ab}=E_{a}-E_{b}$ are energy differences of molecular
states. The coupling coefficient $\hbar\nu_{\text{mol-pl}}=V_{nI=z}$
is identical for all the molecules, cf. Eq.(\ref{eq:molPl}). $k_{b\to f}=k_{b\to f}^{\left(n\right)}$
and $k_{f\to b}=k_{f\to b}^{\left(n\right)}$ are molecular charging
and discharge rate, respectively. If two components of $\mathbf{n}$
change, we indicate the vectors with these components, for example
$\left(n_{gg}-1,n_{ge}+1\right)\equiv\left(n_{gg}-1,n_{ge}+1,n_{gf},n_{eg},n_{ee},n_{ef},n_{fg},n_{fe},n_{ff}\right)$.

We can calculate the population of the product states $\left|\alpha\mu\right\rangle $
with the matrix elements $\rho_{\alpha\mu,\alpha\mu}$. According
to the mapping, these elements correspond to $P_{\left(n_{gg},n_{ee},n_{ff}\right)}^{\mu}\equiv\rho_{\left(n_{gg},0,0,0,n_{ee},0,0,0,n_{ff}\right)}^{\mu,\mu}$.
Here, $n_{gg},n_{ee},n_{ff}$ are the number of molecules, which are in
ground, excited and singly negatively charged state in the molecular
product states $\left|\alpha\right\rangle $, respectively. The population
of the states $\left|\alpha\right\rangle $ is given by $P_{\left(n_{gg},n_{ee},n_{ff}\right)}=\sum_{\mu}P_{\left(n_{gg},n_{ee},n_{ff}\right)}^{\mu}.$
To calculate the population of the plasmon state $\left|\mu\right\rangle $,
we should account for the fact that many $\rho_{\alpha\mu,\beta\nu}$
are mapped to one $\rho_{\mathbf{n}}^{\mu,\nu}$. Finally, we get
$P_{\mu}=\sum_{n_{gg}=0}^{N_{\text{m}}}\sum_{n_{ee}=0}^{N_{\text{m}}-n_{gg}}C_{N_{\text{m}}}^{n_{gg}}C_{N_{\text{m}}-n_{gg}}^{n_{ee}}P_{\left(n_{gg},n_{ee},N_{\text{m}}-n_{gg}-n_{ee}\right)}^{\mu}.$ 

The current through the molecular junction can be calculated with
the reformulated form of Eq. (\ref{eq:current}): $I_{X}=N_{\text{m}}\sum_{a=g,e}\left(k_{Xa\to f}P_{a}-k_{Xf\to a}P_{f}\right).$
The population $P_{nb}=P_{b}$ is identical for all the molecules with the values,
\begin{align}
P_{g} & =\sum_{n_{gg}=1}^{N_{\text{m}}}\sum_{n_{ee}=0}^{N_{\text{m}}-n_{gg}}C_{N_{\text{m}}-1}^{n_{gg}-1}C_{N_{\text{m}}-n_{gg}}^{n_{ee}}P_{\left(n_{gg},n_{ee},N_{\text{m}}-n_{gg}-n_{ee}\right)},
\end{align}
\begin{align}
P_{e} & =\sum_{n_{ee}=1}^{N_{\text{m}}}\sum_{n_{gg}=0}^{N_{\text{m}}-n_{ee}}C_{N_{\text{m}}-1}^{n_{ee}-1}C_{N_{\text{m}}-n_{ee}}^{n_{gg}}P_{\left(n_{gg},n_{ee},N_{\text{m}}-n_{gg}-n_{ee}\right)},
\end{align}
and
\begin{align}
P_{f} & =\sum_{n_{ff}=1}^{N_{\text{m}}}\sum_{n_{ee}=0}^{N_{\text{m}}-n_{ff}}C_{N_{\text{m}}-1}^{n_{ff}-1}C_{N_{\text{m}}-n_{ff}}^{n_{ee}}P_{\left(N_{\text{m}}-n_{ee}-n_{ff},n_{ee},n_{ff}\right)}.
\end{align}

The emission formula given by Eq. (\ref{eq:EmissionOperator}) can
be simplified as follows. Firstly, we notice, cf.
Table \ref{tab:para} and \cite{YZhang}, that the plasmon transition
dipole moment $\mathbf{d}_{I}$ is usually orders of magnitude larger
than the molecular transition dipole moment $\mathbf{d}_{n}$ and, hence, the contribution
containing $\left[\mathbf{d}_{I}\mathbf{d}_{I}^{*}\right]$ dominates
the emission formula. Secondly, only
the plasmon mode $A=B=I=z$ contributes to the system emission (the index $B=z$ will be dropped
in the following). Finally, we get the following expression
\begin{align}
F\left(\omega\right) & =\frac{4\omega^{3}d_{\text{pl}}^{2}}{3\pi c^{3}\hbar}\textrm{Re}\int_{0}^{\infty}dte^{-i\omega t}\sum_{\mu}\sqrt{\mu}\sum_{n_{ff}=0}^{N_{\text{m}}} \sum_{n_{ee}=0}^{N_{\text{m}}-n_{ff}}C_{N_{\text{m}}}^{n_{ff}}C_{N_{\text{m}}-n_{ff}}^{n_{ee}}\nonumber \\
\times & \sigma_{\left(N_{\text{m}}-n_{ff}-n_{ee},0,0,0,n_{ee},0,0,0,n_{ff}\right)}^{\mu-1,\mu}\left(t\right).\label{eq:emsMolJunMNP}
\end{align}
The matrix $\sigma_{\mathbf{n}}^{\mu,\nu}$ satisfies the same set of coupled equations (\ref{eq:molJunRhoN})
as $\rho_{\mathbf{n}}^{\mu,\nu}$, with however the 
initial condition $\sigma_{\mathbf{n}}^{\mu,\nu}(0) = \sqrt{\mu+1}\rho_{\mathbf{n}}^{\mu+1,\nu}\left(\text{ss}\right)$ given by the steady state density matrix.

\subsection{Effect due to Increasing Number of Molecules: Up to 10 Molecules
\label{sec:res_threelevel}}

Simulations of junctions with up to $N_{\rm m}=5$ molecules were presented in \cite{YZhang-1}. Here, the equations (\ref{eq:molJunRhoN}) and
(\ref{eq:emsMolJunMNP}) reduce the computing time and allow calculations up to $N_{\rm m}=10$ molecules, where the number of matrix elements  is reduced from around $3.5\times10^{11}$ of $\rho_{\alpha\mu,\beta\nu}$
to around $4.3\times10^{6}$ of $\rho_{\mathbf{n}}^{\mu,\nu}$.

\begin{figure}
\begin{centering}
\includegraphics[scale=0.5]{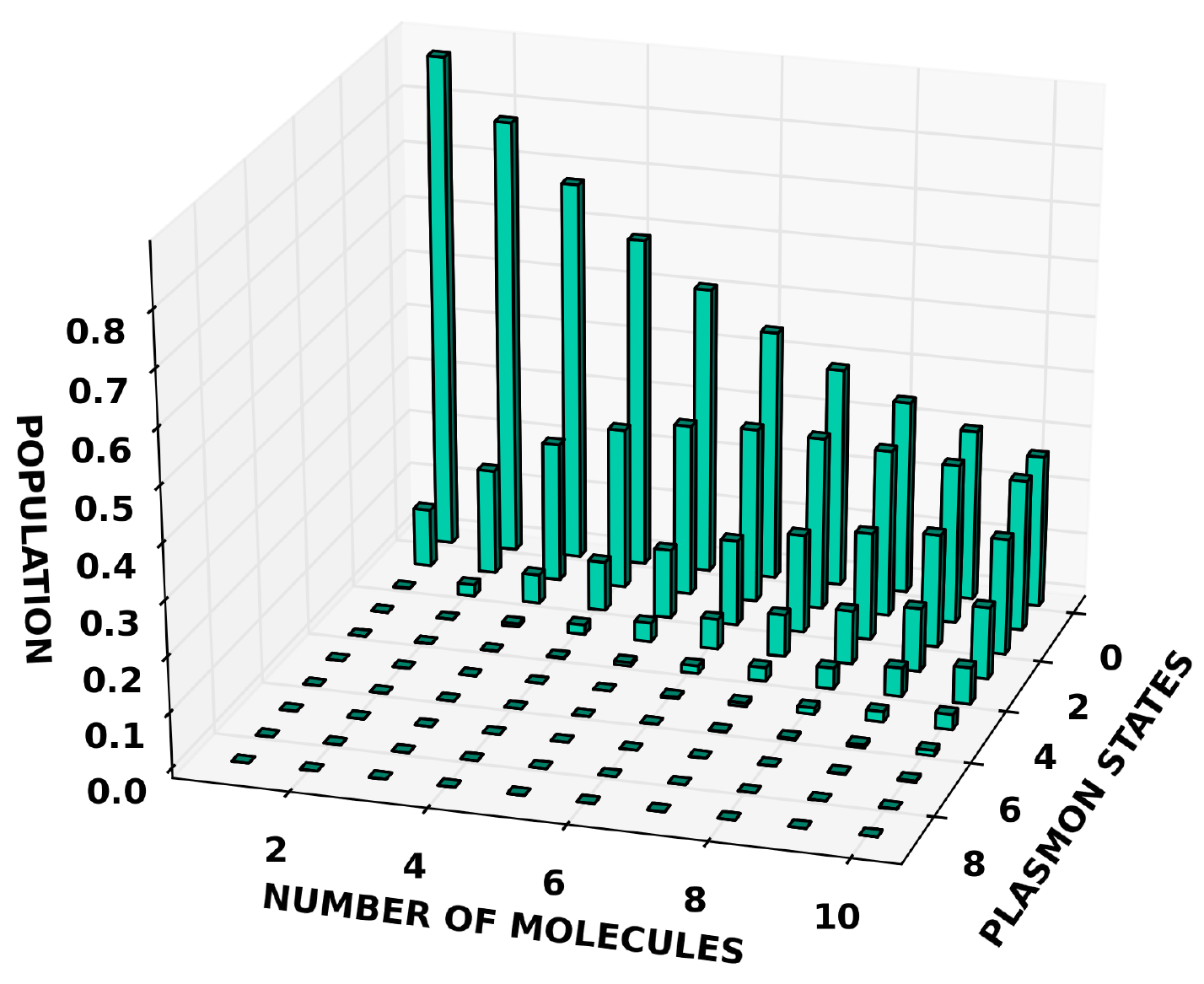}
\par\end{centering}
\begin{centering}
\includegraphics[scale=0.38]{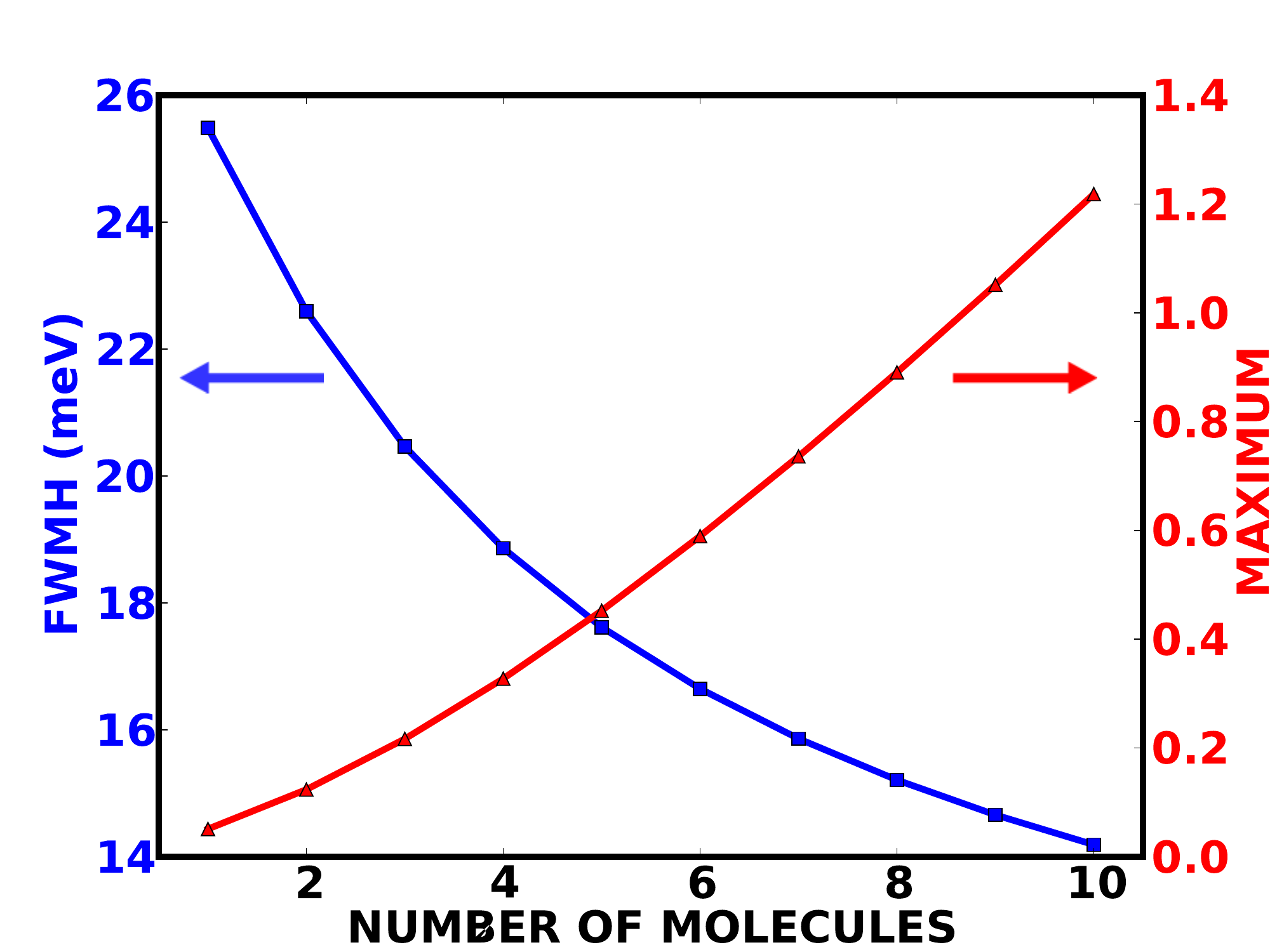}
\par\end{centering}
\begin{centering}
\includegraphics[scale=0.32]{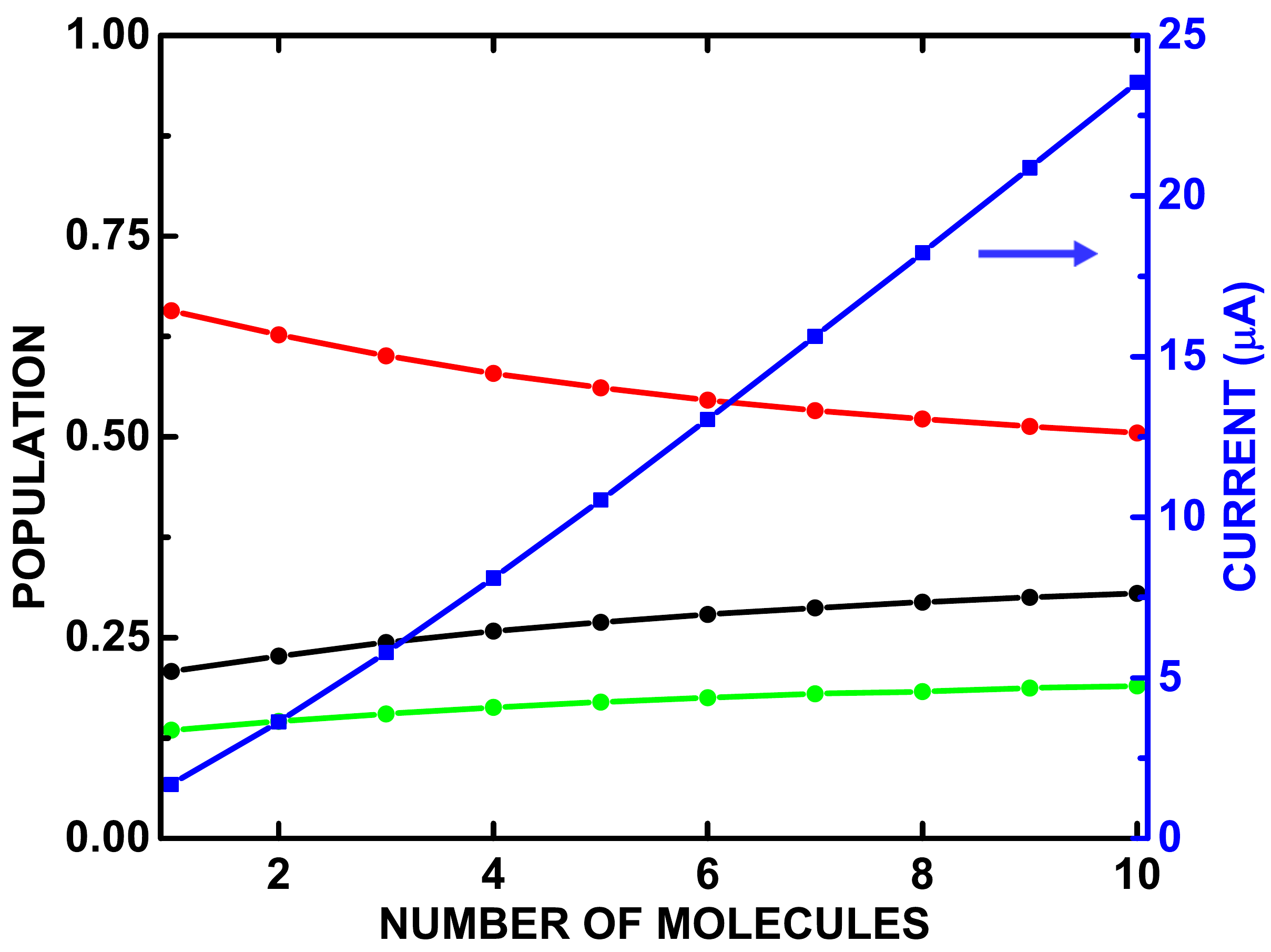}
\par\end{centering}

\caption{\label{fig:molJunMNP}Steady-state properties of junctions with $N_{\text{m}}=1...10$
molecules. Upper panel: population $P_{\mu}$ of excited plasmon states
$\left|\mu\right\rangle $ for different $N_{\text{m}}$. Middle panel,
properties of emission spectra for different $N_{\text{m}}$: blue curve with
squares (left ordinate axis), emission line-width (FWHM); red curve with triangles (right ordinate axis), emission maximum. Lower panel: population of molecular states, red upper curve $P_{e}$,
black middle curve $P_{g}$, and green lower curve $P_{f}$; blue curve (right ordinate axis), current through
the junctions. Other parameters according to Table \ref{tab:para}. }
\end{figure}

The upper panel of Fig.\ref{fig:molJunMNP} shows the population of
plasmon states. Generally speaking, the higher plasmon excited states
are gradually populated when $N_{\mathrm{m}}$ increases, which indicates
the compensation of plasmon damping and an increased plasmon excitation.
The fourth and fifth excited plasmon states are populated for $N_{\text{m}}>5$,
which is in line with the prediction in \cite{YZhang-1}. The middle
panel of Fig.\ref{fig:molJunMNP} shows the maximum and line-width
of the related emission. The emission maximum increases from $0.1$
for $N_{\text{m}}=1$ to $1.2$ for $N_{\text{m}}=10$. 
The spectral line-width reduces from 25 meV for $N_{\text{m}}=1$ to $14$
meV for $N_{\text{m}}=10$. The line narrowing can be explained by the Heisenberg's uncertainty principle \cite{Uncertainty}.  Both of these observations we associate with 
the amplified spontaneous emission of the plasmon. 

The lower panel of Fig.\ref{fig:molJunMNP} shows that the current
through the junction (blue line) increases linearly with the number
of molecules $N_{\text{m}}$. It increases from $1.68$ $\text{\ensuremath{\mu}A}$
for $N_{\text{m}}=1$ to $23.54$ $\text{\ensuremath{\mu}A}$ for
$N_{\text{m}}=10$. The remaining lines indicate the population of
molecular states, which are identical for all the molecules. The population
of the neutral excited state $P_{e}$ (red curve) decreases while the population of the neutral ground state
$P_{g}$ (black curve) and the charged state $P_{f}$ (green curve) increase with increasing
$N_{\text{m}}$ and thus with increasing current through the junction. 

It is expected in \cite{MRichter,YZhang} that when the system
achieves lasing, the population of plasmon number states follows a Poisson-like
distribution. Even with $N_{\rm m}=10$ emitters, we are only approaching the lasing threshold, and while $P_{\mu=1} \gtrsim P_{\mu=0}$, we do not yet see the characteristics of a Poisson distribution.
Therefore, we can not claim the demonstration of lasing, but the calculations give reason to expect a Poisson-like
distribution of plasmon state population for junctions with more molecules.

\section{Approximate Approach Based on Plasmon Reduced Density Matrix\label{sec:approach-plasmon-RDM}}

In this section, an approximate approach motivated by the photon density matrix equation in
the laser theory \cite{Lamb} is proposed for the molecular
junction. A similar approach has been suggested
in \cite{YZhang} to study the plasmonic nano-laser with a molecular optical
pump. The idea is to derive an approximate equation for the plasmon mode
by eliminating the molecular degrees of freedom. The computational
effort by applying this approach is thus only limited by the highest excited
plasmon state involved. Our derivation only assumes that all the molecules couple to one
plasmon mode and although we will apply it to identical molecules it works also for junctions with different molecules.

\subsection{Approximate Equation of Motion for Plasmon Reduced Density Matrix}

The plasmon reduced density matrix is defined by the expectation value:
$\rho_{\mu\nu}\left(t\right)\equiv\text{tr}_{\text{S}}\left\{ \hat{\rho}\left(t\right)\left|\nu\right\rangle \left\langle \mu\right|\right\} $.
The equation of motion for $\rho_{\mu \nu}$ can be obtained with
Eq. (\ref{eq:DensityOperator}):
\begin{eqnarray}
 \frac{\partial}{\partial t}\rho_{\mu\nu}
 & = & -i\omega_{\mu\nu}\rho_{\mu\nu}-\gamma_{\text{pl}}\left[\left(\mu+\nu\right)/2\right]\rho_{\mu\nu}\nonumber \\
 & + & \gamma_{\text{pl}}\sqrt{\left(\mu+1\right)\left(\nu+1\right)}\rho_{\mu+1\nu+1}\nonumber \\
 & - & i\sum_{n=1}^{N_{\rm m}}v_{n}(\sqrt{\mu+1}\rho_{g\mu+1,e\nu}^{\left(n\right)}+\sqrt{\mu}\rho_{e\mu-1,g\nu}^{\left(n\right)}\nonumber \\
 & - & \sqrt{\nu}\rho_{g\mu,e\nu-1}^{\left(n\right)}-\sqrt{\nu+1}\rho_{e\mu,g\nu+1}^{\left(n\right)}).\label{eq:plasmonRDM}
\end{eqnarray}
Here, $\hbar v_{n}$ is the coupling element between the plasmon and
the molecule $n$ and we have introduced $\omega_{\mu\nu}=\left(\mu-\nu\right)\omega_{\text{pl}}$.
This equation depends on the expectation values of two operators:
$\rho_{a\mu,b\nu}^{\left(n\right)}\equiv\text{tr}_{\text{S}}\left\{ \hat{\rho}\left(t\right)\left|b_{n}\right\rangle \left\langle a_{n}\right|\times\left|\nu\right\rangle \left\langle \mu\right|\right\} $,
which describe the correlations of one molecule with the lead plasmon.
The equations of motion for these correlations can be again derived with
Eq. (\ref{eq:DensityOperator}), cf. Appendix \ref{sec:plasmon-rdm-appendix}. As demonstrated there, these equations
depend on the expectation values of three operators, which describe
the correlations of two molecules with the plasmon. Because of the
dissipation the latter correlations are small and thus can be ignored in our treatment to obtain closed equations for $\rho_{\mu\nu}\left(t\right)$, see Eq.
(\ref{eq:SelfConPRDM}).

The diagonal matrix element $\rho_{\mu\mu}$ is
the population $P_{\mu}$ of the plasmon number state $\left|\mu\right\rangle $.
From Eq. (\ref{eq:SelfConPRDM}) we obtain,
\begin{eqnarray}
\frac{\partial}{\partial t}P_{\mu} & = & \left(\gamma_{\text{pl}}\left(\mu+1\right)+k_{\mu+1}\right)P_{\mu+1}-p_{\mu}P_{\mu}\nonumber \\
 & - & \left(\gamma_{\text{pl}}\mu+k_{\mu}\right)P_{\mu}+p_{\mu-1}P_{\mu-1}, \label{eq:plasmon-state-population}
\end{eqnarray}
The rates $k_{\mu}$ due
to the coupling with the molecules reduce the population of higher
plasmon excited states and increase the population of lower ones. In Eq. (\ref{eq:plasmon-state-population}),
the rate $p_{\mu}$ represents the molecule-induced excitation of the plasmon, which has opposite effect compared to $k_{\mu}$ . After some algebra, the molecule-induced plasmon damping and excitation rates can be written as
\begin{equation}
k_{\mu}=\sum_{n=1}^{N_{\rm m}} k_{\mu}^{\left(n\right)}=\mu\sum_{n=1}^{N_{\rm m}}\frac{2\kappa_{\mu}^{\left(n\right)}k_{e\to f}^{\left(n\right)}k_{f\to g}^{\left(n\right)}\Theta^{\left(n\right)}\bar{\Theta}^{\left(n\right)}}{1+2\kappa_{\mu}^{\left(n\right)}\left(\bar{\Theta}^{\left(n\right)}+\bar{\Xi}^{\left(n\right)}\right)\mu},\label{eq:kmu}
\end{equation}
\begin{equation}
p_{\mu-1}=\sum_{n=1}^{N_{\rm m}} p_{\mu-1}^{\left(n\right)}=\mu \sum_{n=1}^{N_{\rm m}} \frac{2\kappa_{\mu}^{\left(n\right)}k_{g\to f}^{\left(n\right)}k_{f\to e}^{\left(n\right)}\Xi^{\left(n\right)}\bar{\Xi}^{\left(n\right)}}{1+2\kappa_{\mu}^{\left(n\right)}\left(\bar{\Theta}^{\left(n\right)}+\bar{\Xi}^{\left(n\right)}\right)\mu}.\label{eq:pmu}
\end{equation}
where $\Theta^{\left(n\right)}=\Theta_{\mu\mu}^{\left(n\right)}$,
$\bar{\Theta}^{\left(n\right)}=\bar{\Theta}_{\mu\mu}^{\left(n\right)}$
as well as $\Xi^{\left(n\right)}=\Xi_{\mu\mu}^{\left(n\right)}$,
$\bar{\Xi}^{\left(n\right)}=\bar{\Xi}_{\mu\mu}^{\left(n\right)}$. These quantities are defined in Eqs. (\ref{eq:theta}) ,(\ref{eq:xi}), (\ref{eq:theta-}) and (\ref{eq:xi-}), and do not depend on the plasmon quantum number $\mu$.  The plasmon state-dependent energy transfer rate is defined as
\begin{equation}
\kappa_{\mu}=\frac{v_{n}^{2}\left(\delta_{n}+\delta_{\mu}\right)}{\left(\omega_{n}-\omega_{\mathrm{pl}}\right)^{2}+\left(\delta_{n}+\delta_{\mu}\right)^{2}}.
\end{equation}
Here, we have introduced the abbreviations: $\hbar\omega_{n}=E_{ne}-E_{ng}$,
$\delta_{n}=\left(k_{g\to f}^{\left(n\right)}+k_{e\to f}^{\left(n\right)}\right)/2$
as well as $\delta_{\mu}=\gamma_{\mathrm{pl}}\left[\left(2\mu-1\right)/2-\sqrt{\mu\left(\mu-1\right)}\right]$.

At steady-state, the time-derivative in Eq. (\ref{eq:plasmon-state-population})
is zero, which leads to a linear algebric equation for $P_{\mu}$:
\begin{eqnarray}
0 & = & \left(\gamma_{\text{pl}}\left(\mu+1\right)+k_{\mu+1}\right)P_{\mu+1}-p_{\mu}P_{\mu}\nonumber \\
 & - & \left(\gamma_{\text{pl}}\mu+k_{\mu}\right)P_{\mu}+p_{\mu-1}P_{\mu-1}.\label{eq:Pmu}
\end{eqnarray}
In the above equation, we get $\left(k_{1}\lambda_{1}+\gamma_{\text{pl}}\right)P_{1}-p_{0}P_{0}=0$
by setting $\mu=0$ (notice $k_{0}=0$ and $p_{-1}=0$). Applying
Eq. (\ref{eq:Pmu}) repeatedly, we can easily get the following recursion
relation
\begin{equation}
r_{\mu}=\frac{P_{\mu}}{P_{\mu-1}}=\frac{p_{\mu-1}}{\gamma_{\text{pl}}\mu+k_{\mu}},\label{eq:recursion-relation}
\end{equation}
which together with normalization $\sum_\mu P_\mu = 1$ readily determines the population distribution of plasmon states. If the ratio $r_{\mu}$ is larger than unity for $\mu < \mu_{c}$ and smaller than
unity for $\mu > \mu_{c}$, the plasmon state population will have a peak-like
distribution around $\mu_{c}$. 

The current through the molecular junction can be evaluated with the
following procedure (for more details, see Appendix \ref{sec:current-appendix}).
According to Eq. (\ref{eq:current}) we should calculate the population
of the molecular states $P_{nc}\equiv\text{tr}_{\text{S}}\left\{ \hat{\rho}\left|c_{n}\right\rangle \left\langle c_{n}\right|\right\} =\sum_{\mu}\rho_{c\mu,c\mu}^{\left(n\right)}$
to determine the current. Obviously, the population is related with
the molecule-plasmon correlations $\rho_{c\mu,c\mu}^{\left(n\right)}$.
These correlations have already been determined when we derive the approximate
equation for the plasmon reduced density matrix $\rho_{\mu\nu}$.
It is shown that they can be connected with $\rho_{\mu\nu}$ and therefore
the current can be determined by $\rho_{\mu\nu}$. If the molecules
are completely identical, the current becomes
\begin{align}
I_{X} & =\left(k_{Xg\to f}+k_{Xf\to g}+k_{Xf\to e}\right)N_{\text{m}}k_{f\to g}k_{e\to f}\Theta\bar{\Theta}\nonumber \\
 & +\left(k_{Xe\to f}+k_{Xf\to g}+k_{Xf\to e}\right)N_{\text{m}}k_{f\to e}k_{g\to f}\Xi\bar{\Xi}\nonumber \\
 & -\left(k_{Xf\to g}+k_{Xf\to e}\right)N_{\text{m}}\nonumber \\
 & +\left(k_{Xg\to f}+k_{Xf\to g}+k_{Xf\to e}\right)\bar{\Theta}\gamma_{\text{pl}}\mathcal{N}_{\text{pl}}\nonumber \\
 & -\left(k_{Xe\to f}+k_{Xf\to g}+k_{Xf\to e}\right)\bar{\Xi}\gamma_{\text{pl}}\mathcal{N}_{\text{pl}}.\label{eq:current-S}
\end{align}
The rates $k_{b\to f}=\sum_{X}k_{Xb\to f}^{\left(n\right)}$ and $k_{f\to b}=\sum_{X}k_{Xf\to b}^{\left(n\right)}$
($b=g,e$) are the total charging and discharging rates while $k_{Xb\to f}=k_{Xb\to f}^{\left(n\right)}$
and $k_{Xf\to b}=k_{Xf\to b}^{\left(n\right)}$ are the parts related
to the specific lead $X$, cf. Eq. (\ref{eq:charging_rate}).
The above formula indicates that the current can be split into two
parts. The first part is contributed by the first three lines in Eq.
(\ref{eq:current-S}) and is proportional to the number of molecules.
This part does not depend on the plasmon excitation. The second part is contributed
by the remaining two lines in Eq. (\ref{eq:current-S}) and is proportional
to the mean number of excited plasmons $\mathcal{N}_{\text{pl}}=\sum_{\mu}\mu P_{\mu}$.

\subsection{Effect due to Increasing Number of Molecules: Up to 50 or more Molecules}

\begin{figure}
\begin{centering}
\includegraphics[scale=0.4]{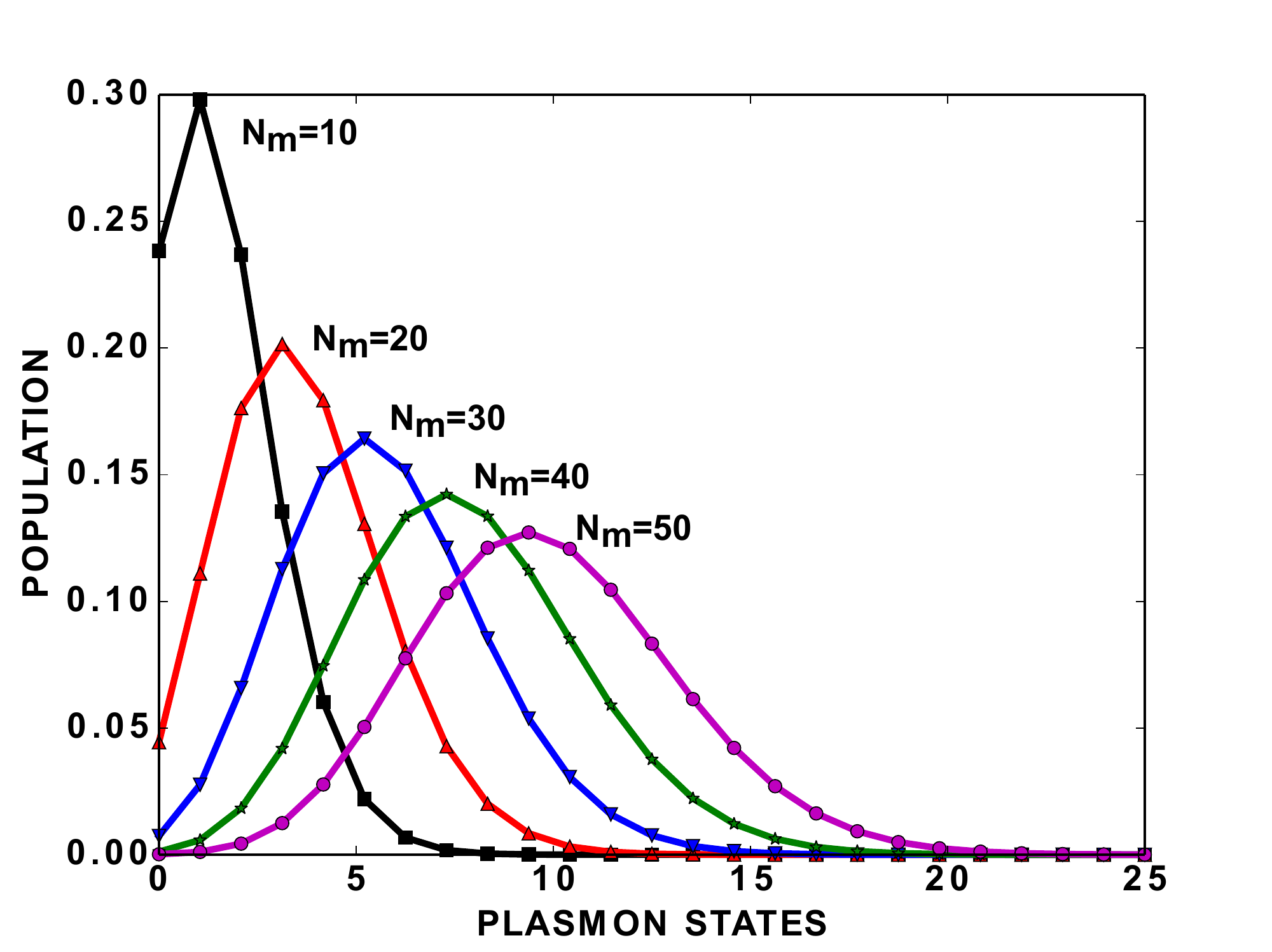}
\par\end{centering}

\begin{centering}
\includegraphics[scale=0.4]{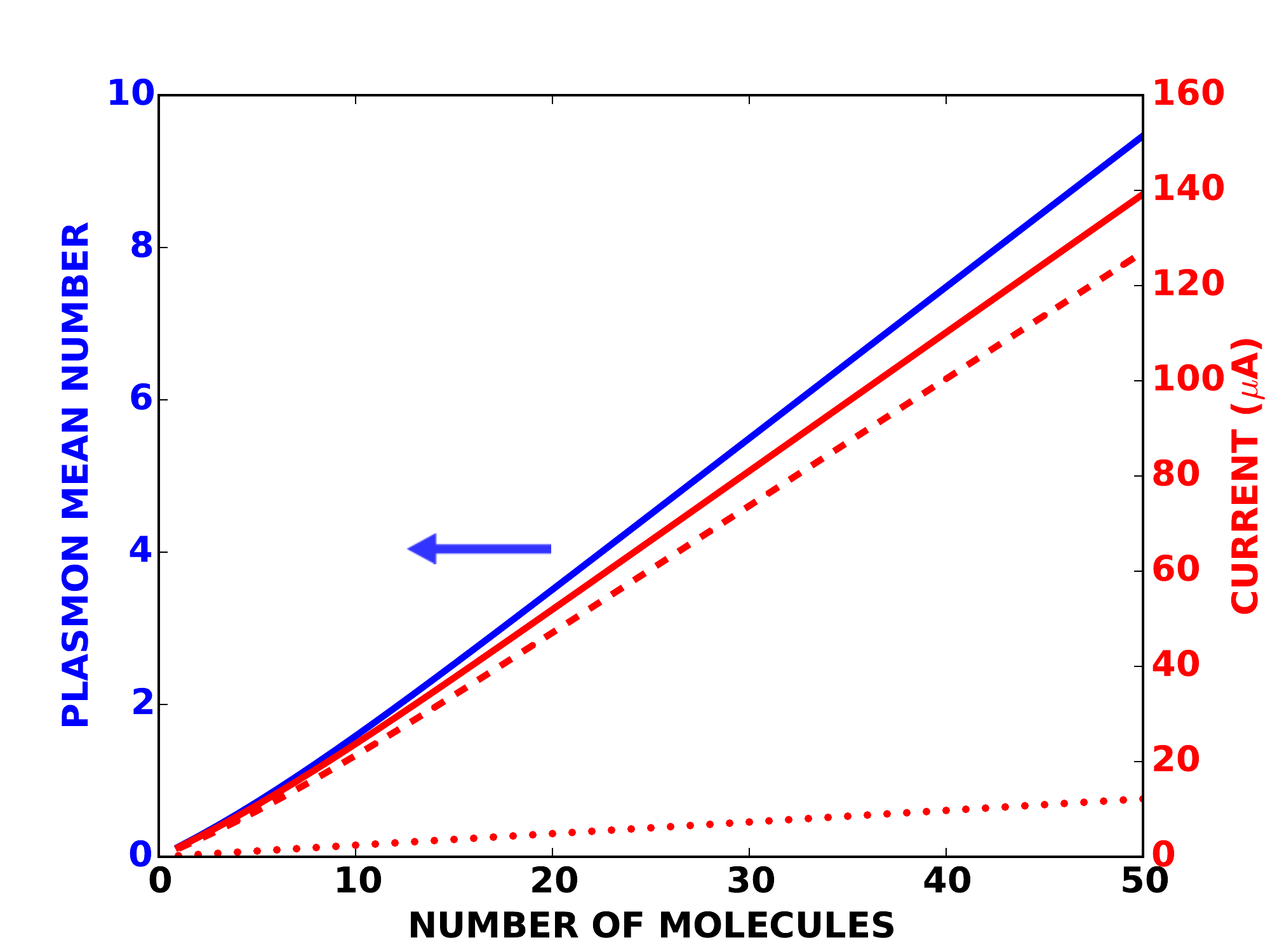}
\par\end{centering}

\caption{\label{fig:recursion}Steady-state properties of junctions with many
molecules. Upper panel: plasmon state population for junctions with
$N_{\text{m}}=10,20,30,40,50$ molecules. Lower panel: blue solid
curve (left ordinate axis), plasmon mean number $\mathcal{N}_{\text{pl}}$; red upper solid curve,
current $I=-I_{L}=I_{R}$ through the junctions; red middle dotted curve, current
contribution proportional to $N_{\text{m}}$, cf. Eq.(\ref{eq:current-S});
red lower dashed curve, current contribution proportional to $\mathcal{N}_{\text{pl}}$,
cf. Eq.(\ref{eq:current-S}). Other parameters according to Table
\ref{tab:para}.}
\end{figure}

We have verified the recursion relation,  Eq. (\ref{eq:recursion-relation}), (thus the approximate
plasmon reduced density matrix) by using it to compute the plasmon
state population for junctions with up to $10$ molecules. The computation
reproduces the exact numerical results, see the upper panel of Fig. \ref{fig:molJunMNP}.
Then, we utilize the recursion relation to compute the plasmon state
population for junctions with up to $50$ molecules, see the upper
panel of Fig.\ref{fig:recursion}. We see that the population distribution
shifts to higher plasmon excited states when $N_{\text{m}}>10$. When
$N_{\text{m}}$ increases from $20$ to $50$, the population distribution
has a peak shape and the peak center is shifted from around $\mu=3$ to around
$\mu=10$. In addition, the plasmon state populations approach the 
Poisson distribution, like the coherent state photon number distribution ascribed to the conventional laser.

We have also calculated the plasmon mean number $\mathcal{N}_{\text{pl}}$
as well as the current through the junctions, cf. lower panel of Fig.\ref{fig:molJunMNP}.
There is almost no excitation for the junction with a single molecule
($\mathcal{N}_{\text{pl}}\approx0$), while we obtain an average of nine plasmon
quanta ($\mathcal{N}_{\text{pl}}=9$) for the junction with $N_{\text{m}}=50$
molecules. The current increases linearly from about $1.6$ $\text{\ensuremath{\mu}A}$
for the junction with single molecule to $140$ $\text{\ensuremath{\mu}A}$
for the junction with $N_{\text{m}}=50$ molecules (cf. red solid
line). The equation (\ref{eq:current-S}) indicates that there are
two contributions to the current. The terms explicitly depending on
$N_{\text{m}}$ are the contribution of the junctions in the absence
of the lead plasmon, cf. the red dotted curve, which is due to electron
transfer processes. The terms explicitly depending on $\mathcal{N}_{\text{pl}}$
are the enhanced current due to the coupling with the lead plasmons, cf.
the red dashed curve. It means extra energy is put into the system through 
the electron transfer process, which is in the end 
utilized to compensate the plasmon damping.

\subsection{Intensity Correlation Function of Emitted Photons}

The emitted photons from the nano-laser shows intensity fluctuations, characterized by the second-order intensity correlation function, which for equal times  
$g^{\left(2\right)}\left(0\right)=\left\langle a^{+}a^{+}aa\right\rangle /\left\langle a^{+}a\right\rangle ^{2}$, is given by a similar expression involving the plasmon mode operators, $g_{\text{pl}}^{\left(2\right)}\left(0\right)\equiv\left\langle C^{+}C^{+}CC\right\rangle /\left\langle C^{+}C\right\rangle ^{2}$. The value thus follows directly from the steady state excitation number distribution calculated above. Photon bunching $(g^{\left(2\right)}\left(0\right) > 1)$, is equivalent to a super Poisson plasmon number distribution with Var$(\mu) > \mathcal{N}_{\text{pl}}=\sum_{\mu}\mu P_{\mu}$.

\begin{figure}
\begin{centering}
\includegraphics[scale=0.55]{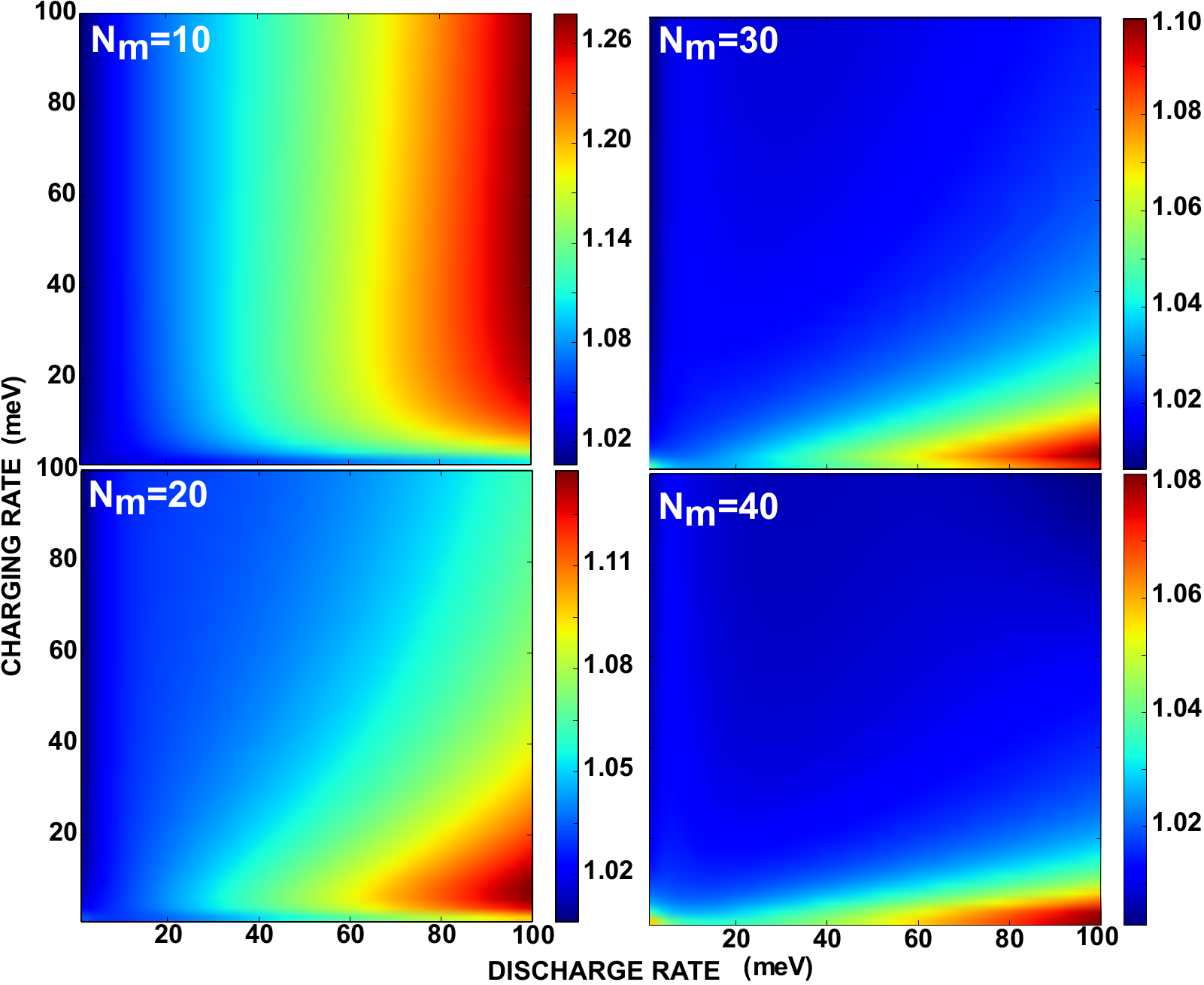}
\par\end{centering}

\caption{\label{fig:correlation}Second order equal-time intensity correlation function $g^{\left(2\right)}\left(0\right)$
for the emitted radiation as a function of the charging rate $\hbar k_{g\to f}^{\left(n\right)}=\hbar\Gamma_{Lgf}^{\left(n\right)}$
and discharging rate $\hbar k_{f\to e}^{\left(n\right)}=\hbar\Gamma_{Rfe}^{\left(n\right)}$
for different values of the number of molecules $N_{\text{m}}$. Other parameters are given in Table \ref{tab:para}.}
\end{figure}

In Fig.\ref{fig:correlation}, the $g_{\text{pl}}^{\left(2\right)}\left(0\right)$
function of junctions with $10, 20, 30$ and $40$ molecules is shown
for different charging $k_{g\to f}^{\left(n\right)}=\Gamma_{Lgf}^{\left(n\right)}$
and discharging rates $k_{f\to e}^{\left(n\right)}=\Gamma_{Rfe}^{\left(n\right)}$.
$g_{\text{pl}}^{\left(2\right)}\left(0\right)$ is always
larger than unity, implying that the emitted photons are bunched and the plasmon number distribution is super-Poisson. 
For a fixed charging rate $k_{g\to f}^{\left(n\right)}$, the bunching increases with increasing discharging rate $k_{f\to e}^{\left(n\right)}$. 
For a fixed $k_{f\to e}^{\left(n\right)}$ it approaches a constant with increasing charging rate $k_{g\to f}^{\left(n\right)}$ for the junction with $10$ molecules, while  junctions with more than $20$ molecules show a decrease towards $g_{\text{pl}}^{\left(2\right)}\left(0\right)=1$ with increasing charging
rate $k_{g\to f}^{\left(n\right)}$, reflecting the approach to Poisson statistics characteristic of lasing. 

\section{Approximate Approach Based on Nonlinear Rate Equations\label{sec:approach-rate-equations} }

In our previous study \cite{YZhang-1} we have derived rate equations
for the molecular state population and plasmon mean number for a junction
with a single molecule. Here, we extend these equations
to junctions with many molecules.

Our starting point is the equations for the population of the molecular states $P_{na}\equiv\text{tr}_{\text{S}}\left\{ \hat{\rho}\left(t\right)\left|a_{n}\right\rangle \left\langle a_{n}\right|\right\} $
(more details see Appendix \ref{sec:rate-equations-appendix}). The equations for $P_{ne}$ and $P_{ng}$ depend
on the molecule-plasmon correlations of the type $\left\langle \left|g_{n}\right\rangle \left\langle e_{n}\right|C^{+}\right\rangle \equiv\text{tr}_{\text{S}}\left\{ \hat{\rho}\left(t\right)\left|g_{n}\right\rangle \left\langle e_{n}\right|C^{+}\right\} $.
These correlations decay much faster than the molecular state populations
because of the plasmon dissipation, and thus we assume that they adiabatically follow the populations $P_{na}$. Inserting the adiabatic solutions of the correlations back into the equations
for $P_{na}$, we get
\begin{equation}
\frac{\partial}{\partial t}P_{nf}=-\left(k_{f\to g}^{\left(n\right)}+k_{f\to e}^{\left(n\right)}\right)P_{nf}+k_{e\to f}^{\left(n\right)}P_{ne}+k_{g\to f}^{\left(n\right)}P_{ng},\label{eq:pnf}
\end{equation}
\begin{equation}
\frac{\partial}{\partial t}P_{ne}=-k_{e\to f}^{\left(n\right)}P_{ne}+k_{f\to e}^{\left(n\right)}P_{nf}+\kappa_{n}\mathcal{N}_{\text{pl}}P_{ng}-\kappa_{n}\left[1+\mathcal{N}_{\text{pl}}\right]P_{ne},\label{eq:pneF}
\end{equation}
\begin{equation}
\frac{\partial}{\partial t}P_{ng}=-k_{g\to f}^{\left(n\right)}P_{ng}+k_{f\to g}^{\left(n\right)}P_{nf}-\kappa_{n}\mathcal{N}_{\text{pl}}P_{ng}+\kappa_{n}\left[1+\mathcal{N}_{\text{pl}}\right]P_{ne}.\label{eq:pngF}
\end{equation}
Here, the energy transfer rates are defined as $\kappa_{n}=2v_{n}^{2}\gamma_{n}/\left[\left(\omega_{\text{pl}}-\omega_{n}\right)^{2}+\gamma_{n}^{2}\right]$
with $\gamma_{n}=\left(\gamma_{\text{pl}}+k_{e\to f}^{\left(n\right)}+k_{g\to f}^{\left(n\right)}\right)/2$.
Notice the above equations depend on the plasmon mean number $\mathcal{N}_{\text{pl}}\equiv\left\langle C^{+}C\right\rangle $.
The equation for $\mathcal{N}_{\text{pl}}$ also depends on the molecule-plasmon
correlations. With the adiabatic solution for the correlations, we can get the following equation 
\begin{equation}
\frac{\partial}{\partial t} \mathcal{N}_{\text{pl}}=\sum_{n=1}^{N_{\rm m}}\kappa_{n}\left(P_{ne}+ \mathcal{N}_{\text{pl}}\left(P_{ne}-P_{ng}\right)\right)-\gamma_{\text{pl}}\mathcal{N}_{\text{pl}},\label{eq:Npl}
\end{equation}
where $\kappa_{n}P_{ne}$, $\kappa_{n}\mathcal{N}_{\text{pl}}P_{ne}$ and $-\kappa_{n}\mathcal{N}_{\text{pl}}P_{ng}$ represent spontaneous and stimulated emission, as well as  stimulated absorption of plasmon excitation by the molecules.

\subsection{Comparison of the Plasmon Mean Number Calculated with Different Approaches}

\begin{figure}
\begin{centering}
\includegraphics[scale=0.4]{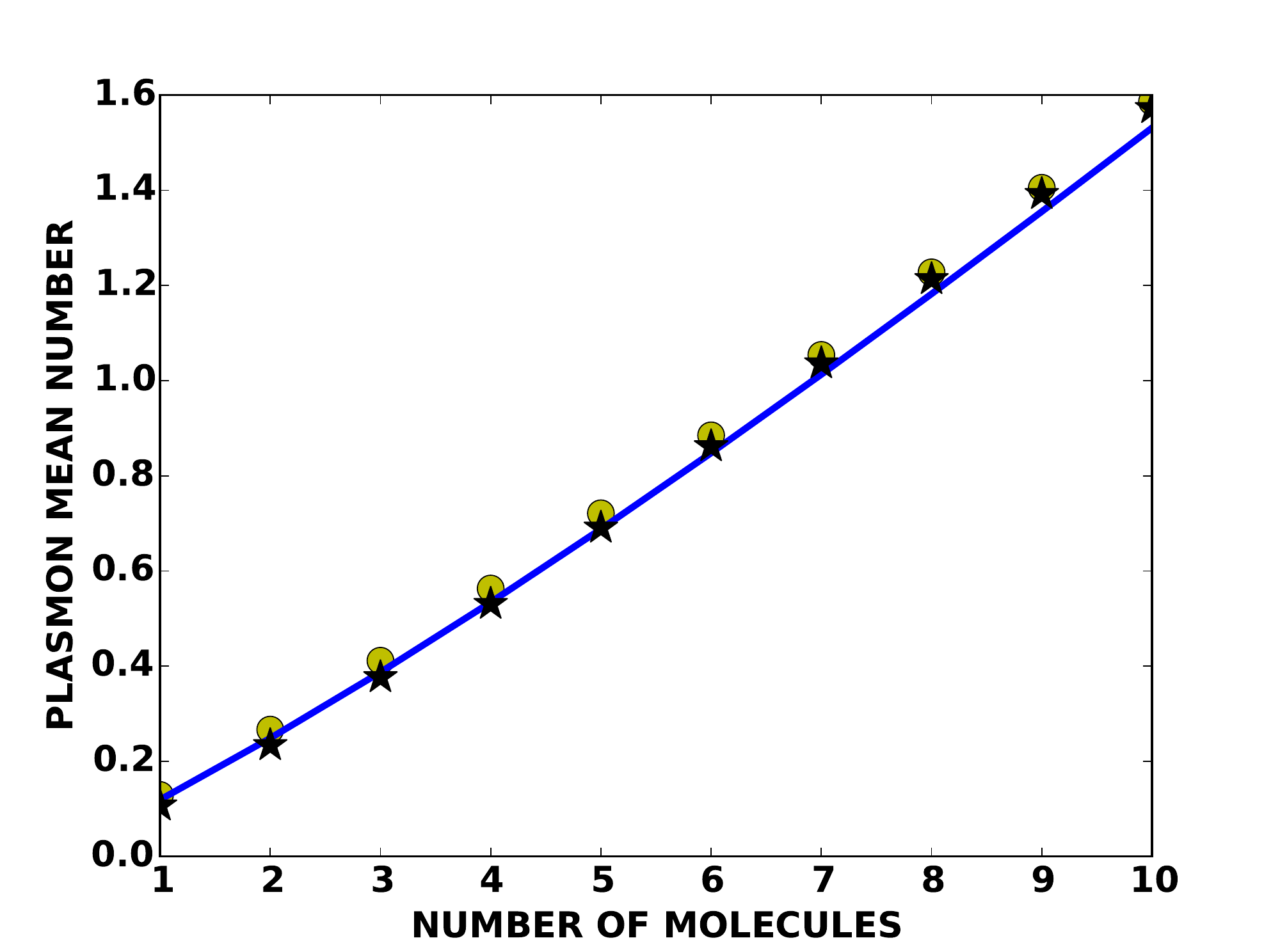}
\par\end{centering}

\caption{\label{fig:MeanNumberRate}Plasmon mean number calculated with different
approaches for junctions with different numbers of molecules $N_{\text{m}}$.
Black stars: calculations based on simplified reduced density matrix
equation Eq.(\ref{eq:molJunRhoN}). Yellow circles: calculations based
on recursion relation Eq.(\ref{eq:recursion-relation}). Blue solid
line: calculations based on rate equations (\ref{eq:pnf}),(\ref{eq:pneF}),(\ref{eq:pngF})
and (\ref{eq:Npl}). Other parameters according to Table \ref{tab:para}.}
\end{figure}

In Fig.\ref{fig:MeanNumberRate}, we compare the different approaches
for the calculation of the plasmon mean number for junctions with different
numbers of molecules. These approaches are based on the full
reduced density matrix (RDM) equation (black stars), the population
recursion relation (derived from the plasmon RDM, yellow circles) as well
as the rate equations (blue solid line). The plasmon mean number increases
almost linearly with increasing number of molecules. The green squares
approach the black dots from above when the number of molecules in the junction increases and this proves the validity
of the plasmon RDM in that limit. We have also compared the results of the  plasmon RDM  with the results of the rate equations for junctions with up to $50$ molecules and found that
the latter slightly underestimate the plasmon mean number (not shown). We
explain this by the overestimation of the spontaneous energy transfer,
cf. the discussion about Fig. 7 in \cite{YZhang}.

\section{Conclusions\label{sec:conclusions}}

We have extended the study \cite{YZhang-1} of the electroluminescence of a molecular junction
excited through an energy exchange coupling with electron-transfer induced excited molecules to the junctions with many molecules. In the present article, we simplified the full system master equation by utilizing symmetries of the density matrix for identical molecules. We carried out exact simulations for junctions with up to
$10$ molecules. With increasing number of molecules, higher excited
states of the lead plasmon are populated, accompanied by a
narrowing of the emission, indicating the amplified emission of the plasmon.

Our analysis did not incorporate the coherence induced by the
excitonic and charge transfer coupling among the molecules. 
The excitonic coupling between molecules can in principle 
be introduced in the system Hamiltonian, while the charge
transfer coupling may be treated as incoherent terms in the RDM equations.
These couplings become important if the molecular ensemble is dense and constitute an intersting topic for further studies.

Approximate equations of motion for the plasmon degrees of freedom were derived, and for junctions
in steady-state, a recursion relation was obtained for the plasmon state
populations. The population distribution of the plasmon states is
Poisson-like, and the intensity fluctuations of the emitted radiation are reduced for junctions with more than $20$ molecules, which indicates
the formation of a plasmon coherent state. 
Finally, non-linear rate equations were derived for the molecular state populations
and the average plasmon excitation, which well account for the main features found by the other methods.

Thus validated, the symmetric approaches
may also be applied to junctions with many different molecules and 
to junctions, where the molecules couple to several plasmon modes. Our analysis is 
carried out for a  particular pumping mechanism involving electron transfer to excited molecular states, but the 
approaches outlined may find wider application for other systems with similar excitation mechanisms, for example, the 
optically pumped nano-laser \cite{YZhang-5} and the semiconductor-based plasmonic nano-laser \cite{KDing,MTHill}.

\begin{acknowledgments}
Y.Z. ackowledges Yaroslav Zelinskyy, Dirk Ziemann and Thomas
Plehn for several illuminating discussions. This work was supported by the \emph{Deutsche Forschungsgemeinschaft }through
Sfb 951 and by the GIF research Grant No. 1146-73.14/2011 (V. M.)
, by the\emph{ China Scholarship Council }(Y. Z.), as well as the Villum
Foundation (Y. Z and K. M.).
\end{acknowledgments}

\appendix

\section{Derivation of the Approximate Plasmon Reduced Density Matrix Equation\label{sec:plasmon-rdm-appendix} }
The equations for the molecule-plasmon correlations $\rho_{a\mu,b\nu}^{\left(n\right)}$ can be written as
\begin{eqnarray}
 &  & \frac{\partial}{\partial t}\rho_{g\mu,g\nu}^{\left(n\right)}\nonumber \\
 & = & -i\omega_{\mu\nu}\rho_{g\mu,g\nu}^{\left(n\right)}-k_{g\to f}^{\left(n\right)}\rho_{g\mu,g\nu}^{\left(n\right)}+k_{f\to g}^{\left(n\right)}\rho_{f\mu,f\nu}^{\left(n\right)}\nonumber \\
 & - & \gamma_{\text{pl}}\left[\left(\mu+\nu\right)/2\right]\rho_{g\mu,g\nu}^{\left(n\right)}+\gamma_{\text{pl}}\sqrt{\left(\mu+1\right)\left(\nu+1\right)}\rho_{g\mu+1,g\nu+1}^{\left(n\right)}\nonumber \\
 & + & iv_{n}\left(\sqrt{\nu}\rho_{g\mu,e\nu-1}^{\left(n\right)}-\sqrt{\mu}\rho_{e\mu-1,g\nu}^{\left(n\right)}\right),\label{eq:rhogg}
\end{eqnarray}
\begin{eqnarray}
 &  & \frac{\partial}{\partial t}\rho_{g\mu,e\nu-1}^{\left(n\right)}\nonumber \\
 & = & i\left(\tilde{\omega}_{n}^*-\omega_{\mu\nu-1}\right)\rho_{g\mu,e\nu-1}^{\left(n\right)}\nonumber \\
 & - & \gamma_{\text{pl}}\left[\left(\mu+\nu-1\right)/2\right]\rho_{g\mu,e\nu-1}^{\left(n\right)}+\gamma_{\text{pl}}\sqrt{\left(\mu+1\right)\nu}\rho_{g\mu+1,e\nu}^{\left(n\right)}\nonumber \\
 & + & iv_{n}\left(\sqrt{\nu}\rho_{g\mu,g\nu}^{\left(n\right)}-\sqrt{\mu}\rho_{e\mu-1,e\nu-1}^{\left(n\right)}\right),\label{eq:rhoge}
\end{eqnarray}
\begin{eqnarray}
 &  & \frac{\partial}{\partial t}\rho_{e\mu-1,g\nu}^{\left(n\right)}\nonumber \\
 & = & -i\left(\tilde{\omega}_{n}+\omega_{\mu-1\nu}\right)\rho_{e\mu-1,g\nu}^{\left(n\right)} \nonumber \\
 & - & \gamma_{\text{pl}}\left[\left(\mu-1+\nu\right)/2\right]\rho_{e\mu-1,g\nu}^{\left(n\right)}+\gamma_{\text{pl}}\sqrt{\mu\left(\nu+1\right)}\rho_{e\mu,g\nu+1}^{\left(n\right)}\nonumber \\
 & + & iv_{n}\left(\sqrt{\nu}\rho_{e\mu-1,e\nu-1}^{\left(n\right)}-\sqrt{\mu}\rho_{g\mu,g\nu}^{\left(n\right)}\right),\label{eq:rhoeg}
\end{eqnarray}
\begin{eqnarray}
 &  & \frac{\partial}{\partial t}\rho_{e\mu-1,e\nu-1}^{\left(n\right)}\nonumber \\
 & = & -i\omega_{\mu-1\nu-1}\rho_{e\mu-1,e\nu-1}^{\left(n\right)}-k_{e\to f}^{\left(n\right)}\rho_{e\mu-1,e\nu-1}^{\left(n\right)}+k_{f\to e}^{\left(n\right)}\rho_{f\mu-1,f\nu-1}^{\left(n\right)}\nonumber \\
 & - & \gamma_{\text{pl}}\left[\left(\mu+\nu-2\right)/2\right]\rho_{e\mu-1,e\nu-1}^{\left(n\right)}+\gamma_{\text{pl}}\sqrt{\mu\nu}\rho_{e\mu,e\nu}^{\left(n\right)}\nonumber \\
 & + & iv_{n}\left(\sqrt{\nu}\rho_{e\mu-1,g\nu}^{\left(n\right)}-\sqrt{\mu}\rho_{g\mu,e\nu-1}^{\left(n\right)}\right),\label{eq:rhoee}
\end{eqnarray}
\begin{eqnarray}
 &  & \frac{\partial}{\partial t}\rho_{f\mu,f\nu}^{\left(n\right)}\nonumber \\
 & = & -i\omega_{\mu\nu}\rho_{f\mu,f\nu}^{\left(n\right)}+k_{g\to f}^{\left(n\right)}\rho_{g\mu,g\nu}^{\left(n\right)}+k_{e\to f}^{\left(n\right)}\rho_{e\mu,e\nu}^{\left(n\right)}\nonumber \\
 & - & \gamma_{\text{pl}}\left[\left(\mu+\nu\right)/2\right]\rho_{f\mu-1,f\nu-1}^{\left(n\right)}+\gamma_{\text{pl}}\sqrt{\left(\mu+1\right)\left(\nu+1\right)}\rho_{f\mu,f\nu}^{\left(n\right)}\nonumber \\
 & - & \left(k_{f\to g}^{\left(n\right)}+k_{f\to e}^{\left(n\right)}\right)\rho_{f\mu,f\nu}^{\left(n\right)}.\label{eq:rhoff}
\end{eqnarray}
Here, we have introduced the complex transition frequencies  $\tilde{\omega}_{n}=\omega_{n}-i\delta_{n}$
with $\delta_{n}=\left(k_{g\to f}^{\left(n\right)}+k_{e\to f}^{\left(n\right)}\right)/2$. The above equations actually also depend on the correlations $\rho_{ab\mu,cd\nu}^{\left(n,n'\right)}\equiv\text{tr}_{\text{S}}\left\{ \hat{\rho}\left(t\right)\left|c_{n}\right\rangle \left\langle a_{n}\right|\times\left|d_{n'}\right\rangle \left\langle b_{n'}\right|\times\left|\nu\right\rangle \left\langle \mu\right|\right\} $ ($n\neq n'$) of two different molecules and the plasmon mode. These coherences decay faster than the single molecule-plasmon correlations $\rho_{a\mu,b\nu}^{\left(n\right)}$, and therefore are neglected in our approximate treatment. Next, we assume a moderate variation of density matrix elements with the plasmon number and replace the term 
$\sqrt{\left(\mu+1\right)\left(\nu+1\right)}\rho_{a\mu+1,b\nu+1}^{\left(n\right)}$ by $\sqrt{\mu\nu}\rho_{a\mu,b\nu}^{\left(n\right)}$, and write
\begin{eqnarray}
 & - & \left[i\omega_{\mu\nu}-\gamma_{\text{pl}}\left(\mu+\nu\right)/2\right]\rho_{a\mu,b\nu}^{\left(n\right)}\nonumber \\
 & + & \gamma_{\text{pl}}\sqrt{\left(\mu+1\right)\left(\nu+1\right)}\rho_{a\mu+1,b\nu+1}^{\left(n\right)}\to-i\tilde{\omega}_{\mu\nu}\rho_{a\mu,b\nu}^{\left(n\right)},
\end{eqnarray}
where we have introduced the complex frequency $\tilde{\omega}_{\mu\nu}=\omega_{\mu\nu}-i\gamma_{\text{pl}}\left[\left(\mu+\nu\right)/2-\sqrt{\mu\nu}\right]$.

To proceed we notice that only the dissipation of
the plasmon contributes to the equation (\ref{eq:plasmonRDM}) for
$\rho_{\mu\nu}$. In contrast, the dissipation of both the molecule
and the plasmon contributes to the equations for $\rho_{a\mu,b\nu}^{\left(n\right)}$.
This implies that $\rho_{a\mu,b\nu}^{\left(n\right)}$ change much faster and thus may {\it adiabatically}
follow the change of $\rho_{\mu\nu}$. The approximate version of the equations (\ref{eq:rhogg}) to (\ref{eq:rhoff}) do not explicitly depend on $\rho_{\mu\nu}$. However, due to the
relation $\rho_{f\mu,f\nu}^{\left(n\right)}+\rho_{e\mu e\nu}^{\left(n\right)}+\rho_{g\mu g\nu}^{\left(n\right)}=\rho_{\mu\nu}$, they actually implicitly couple with the equation (\ref{eq:plasmonRDM})
for $\rho_{\mu\nu}$. From Eq. (\ref{eq:rhoff}), we express $\rho_{f \mu,f \nu} ^{(n)}$ with $\rho_{e \mu,e \nu} ^{(n)}$ by replacing $\rho_{g \mu,g \nu} ^{(n)}$ with $\rho_{\mu,\nu} - \rho_{f \mu,f \nu} ^{(n)} -\rho_{e \mu,e \nu} ^{(n)}$. Similarly, we can also express $\rho_{f \mu,f \nu} ^{(n)}$ with $\rho_{g \mu,g \nu} ^{(n)}$ by replacing $\rho_{e \mu,e \nu} ^{(n)}$ with $\rho_{\mu,\nu} - \rho_{f \mu,f \nu} ^{(n)} -\rho_{g \mu,g \nu} ^{(n)}$.
Then, we insert the results to Eqs. (\ref{eq:rhoge}) and (\ref{eq:rhoeg}) and get 

\begin{eqnarray}
\rho_{g\mu,e\nu-1}^{\left(n\right)} & = & \lambda_{\mu\nu}^{\left(n\right)}\left(a_{\mu\nu}^{\left(n\right)}\sqrt{\mu}+c_{\mu\nu}^{\left(n\right)}\sqrt{\nu}\right)g_{\mu-1\nu-1}^{\left(n\right)}\rho_{\mu-1\nu-1}\nonumber \\
 & - & \lambda_{\mu\nu}^{\left(n\right)}\left(a_{\mu\nu}^{\left(n\right)}\sqrt{\nu}+c_{\mu\nu}^{\left(n\right)}\sqrt{\mu}\right)f_{\mu\nu}^{\left(n\right)}\rho_{\mu\nu},\label{eq:rhogeS}
\end{eqnarray}
\begin{eqnarray}
\rho_{e\mu-1,g\nu}^{\left(n\right)} & = & \lambda_{\mu\nu}^{\left(n\right)}\left(b_{\mu\nu}^{\left(n\right)}\sqrt{\nu}-c_{\mu\nu}^{\left(n\right)}\sqrt{\mu}\right)g_{\mu-1\nu-1}^{\left(n\right)}\rho_{\mu-1\nu-1}\nonumber \\
 & - & \lambda_{\mu\nu}^{\left(n\right)}\left(b_{\mu\nu}^{\left(n\right)}\sqrt{\mu}-c_{\mu\nu}^{\left(n\right)}\sqrt{\nu}\right)f_{\mu\nu}^{\left(n\right)}\rho_{\mu\nu}.\label{eq:rhoegS}
\end{eqnarray}
In the above expressions, we have introduced the following abbreviations
\begin{align}
a_{\mu\nu}^{\left(n\right)} & =\left(\tilde{\omega}_{n}+\tilde{\omega}_{\mu-1\nu}\right)-iv_{n}^{2}\left(\bar{\Xi}_{\mu-1\nu-1}^{\left(n\right)}\nu+\bar{\Theta}_{\mu\nu}^{\left(n\right)}\mu\right),\\
b_{\mu\nu}^{\left(n\right)} & =\left(\tilde{\omega}_{n}^{*}-\tilde{\omega}_{\mu\nu-1}\right)+iv_{n}^{2}\left(\bar{\Theta}_{\mu\nu}^{\left(n\right)}\nu+\bar{\Xi}_{\mu-1\nu-1}^{\left(n\right)}\mu\right),\\
c_{\mu\nu}^{\left(n\right)} & =iv_{n}^{2}\left(\bar{\Theta}_{\mu\nu}^{\left(n\right)}+\bar{\Xi}_{\mu-1\nu-1}^{\left(n\right)}\right)\sqrt{\mu\nu},\\
f_{\mu\nu}^{\left(n\right)} & =v_{n}k_{f\to g}^{\left(n\right)}k_{e\to f}^{\left(n\right)}\Theta_{\mu\nu}^{\left(n\right)}\bar{\Theta}_{\mu\nu}^{\left(n\right)},\\
g_{\mu\nu}^{\left(n\right)} & =v_{n}k_{f\to e}^{\left(n\right)}k_{g\to f}^{\left(n\right)}\Xi_{\mu\nu}^{\left(n\right)}\bar{\Xi}_{\mu\nu}^{\left(n\right)},
\end{align}
as well as
\begin{equation}
1/\lambda_{\mu\nu}^{\left(n\right)}=a_{\mu\nu}^{\left(n\right)}b_{\mu\nu}^{\left(n\right)}+c_{\mu\nu}^{\left(n\right)}c_{\mu\nu}^{\left(n\right)}.
\end{equation}
In addition, the following abbreviations have been used:
\begin{eqnarray}
1/\Theta_{\mu\nu}^{\left(n\right)} & = & i\tilde{\omega}_{\mu\nu}+k_{f\to g}^{\left(n\right)}+k_{f\to e}^{\left(n\right)}+k_{e\to f}^{\left(n\right)},\label{eq:theta}\\
1/\Xi_{\mu\nu}^{\left(n\right)} & = & i\tilde{\omega}_{\mu\nu}+k_{f\to g}^{\left(n\right)}+k_{f\to e}^{\left(n\right)}+k_{g\to f}^{\left(n\right)},\label{eq:xi}\\
1/\bar{\Theta}_{\mu\nu}^{\left(n\right)} & = & i\tilde{\omega}_{\mu\nu}+k_{g\to f}^{\left(n\right)}-k_{f\to g}^{\left(n\right)}\left(k_{g\to f}^{\left(n\right)}-k_{e\to f}^{\left(n\right)}\right)\Theta_{\mu\nu}^{\left(n\right)},\label{eq:theta-}\\
1/\bar{\Xi}_{\mu\nu}^{\left(n\right)} & = & i\tilde{\omega}_{\mu\nu}+k_{e\to f}^{\left(n\right)}-k_{f\to e}^{\left(n\right)}\left(k_{e\to f}^{\left(n\right)}-k_{g\to f}^{\left(n\right)}\right)\Xi_{\mu\nu}^{\left(n\right)}.\label{eq:xi-}
\end{eqnarray}
The solutions for $\rho_{g\mu,g\nu}^{\left(n\right)}$ and $\rho_{e\mu-1,e\nu-1}^{\left(n\right)}$
are
\begin{eqnarray}
\rho_{g\mu,g\nu}^{\left(n\right)} & = & iv_{n}\bar{\Theta}_{\mu\nu}^{\left(n\right)}(i_{\mu\nu}^{\left(n\right)}\rho_{\mu\nu}-h_{\mu\nu}^{\left(n\right)}g_{\mu-1\nu-1}^{\left(n\right)}\rho_{\mu-1\nu-1})\nonumber \\
 & + & k_{f\to g}^{\left(n\right)}k_{e\to f}^{\left(n\right)}\Theta_{\mu\nu}^{\left(n\right)}\bar{\Theta}_{\mu\nu}^{\left(n\right)}\rho_{\mu\nu},\label{eq:rhoggS}
\end{eqnarray}
\begin{eqnarray}
\rho_{e\mu-1,e\nu-1}^{\left(n\right)} & = & iv_{n}\bar{\Xi}_{\mu-1\nu-1}^{\left(n\right)}(j_{\mu\nu}^{\left(n\right)}g_{\mu-1\nu-1}^{\left(n\right)}\rho_{\mu-1\nu-1}-h_{\mu\nu}^{\left(n\right)}f_{\mu\nu}^{\left(n\right)}\rho_{\mu\nu})\nonumber \\
 & + & k_{f\to e}^{\left(n\right)}k_{g\to f}^{\left(n\right)}\Xi_{\mu-1\nu-1}^{\left(n\right)}\bar{\Xi}_{\mu-1\nu-1}^{\left(n\right)}\rho_{\mu-1\nu-1},\label{eq:rhoeeS}
\end{eqnarray}
where we have introduced the following abbreviations:
\begin{align}
h_{\mu\nu}^{\left(n\right)} & =\lambda_{\mu\nu}^{\left(n\right)}\left[\left(b_{\mu\nu}^{\left(n\right)}-a_{\mu\nu}^{\left(n\right)}\right)\sqrt{\mu\nu}-c_{\mu\nu}^{\left(n\right)}\left(\mu+\nu\right)\right],\\
i_{\mu\nu}^{\left(n\right)} & =\lambda_{\mu\nu}^{\left(n\right)}\left(b_{\mu\nu}^{\left(n\right)}\mu-a_{\mu\nu}^{\left(n\right)}\nu-2c_{\mu\nu}^{\left(n\right)}\sqrt{\mu\nu}\right),\\
j_{\mu\nu}^{\left(n\right)} & =\lambda_{\mu\nu}^{\left(n\right)}\left(b_{\mu\nu}^{\left(n\right)}\nu-a_{\mu\nu}^{\left(n\right)}\mu-2c_{\mu\nu}^{\left(n\right)}\sqrt{\mu\nu}\right).
\end{align}
Inserting Eqs. (\ref{eq:rhogeS}) and (\ref{eq:rhoegS}) into Eq.
(\ref{eq:plasmonRDM}), we finally get the master equation for the plasmon reduced density matrix
\begin{eqnarray}
\frac{\partial}{\partial t}\rho_{\mu\nu} & = & -i\omega_{\mu\nu}\rho_{\mu\nu}-\gamma_{\text{pl}}\left[\left(\mu+\nu\right)/2\right]\rho_{\mu\nu}\nonumber \\
 & + & \gamma_{\text{pl}}\sqrt{\left(\mu+1\right)\left(\nu+1\right)}\rho_{\mu+1\nu+1}\nonumber \\
 & - & i\sum_{n}v_{n}(h_{\mu+1\nu+1}^{\left(n\right)}f_{\mu+1\nu+1}^{\left(n\right)}\rho_{\mu+1\nu+1}\nonumber \\
 & + & h_{\mu\nu}^{\left(n\right)}g_{\mu-1\nu-1}^{\left(n\right)}\rho_{\mu-1\nu-1}\nonumber \\
 & - & j_{\mu+1\nu+1}^{\left(n\right)}g_{\mu\nu}^{\left(n\right)}\rho_{\mu\nu}-i_{\mu\nu}^{\left(n\right)}f_{\mu\nu}^{\left(n\right)}\rho_{\mu\nu}).\label{eq:SelfConPRDM}
\end{eqnarray}

The equation for the populations $\rho_{\mu\mu}$  is given by Eq. (\ref{eq:plasmon-state-population})
in the main text. In that equation, the rates induced by the molecules
are defined as $k_{\mu}\equiv-i\sum_{n}v_{n}h_{\mu\mu}^{\left(n\right)}f_{\mu\mu}^{\left(n\right)}$
and $p_{\mu}\equiv-i\sum_{n}v_{n}h_{\mu+1\mu+1}^{\left(n\right)}g_{\mu\mu}^{\left(n\right)}$.

\section{Current Through the Molecular Junction\label{sec:current-appendix}}

The current through the molecular junction is be calculated with
the formula (\ref{eq:current}). That formula depends on the populations of molecular levels
$P_{nb}\equiv\text{tr}\left\{ \hat{\rho}\left|b_{n}\right\rangle \left\langle b_{n}\right|\right\} =\sum_{\mu}\rho_{b\mu,b\mu}^{\left(n\right)}$. From Eqs. (\ref{eq:rhoggS}) and (\ref{eq:rhoeeS}), we directly get 
the expressions for the molecule-plasmon correlations $\rho_{g\mu,g\mu}^{\left(n\right)}$
and $\rho_{e\mu,e\mu}^{\left(n\right)}$ :
\begin{eqnarray}
\rho_{g\mu,g\mu}^{\left(n\right)} & = & \bar{\Theta}^{\left(n\right)}\left(p_{\mu-1}^{\left(n\right)}P_{\mu-1}-k_{\mu}^{\left(n\right)}P_{\mu}\right)\nonumber \\
 & + & k_{f\to g}^{\left(n\right)}k_{e\to f}^{\left(n\right)}\Theta^{\left(n\right)}\bar{\Theta}^{\left(n\right)}P_{\mu},\label{eq:rhoggSS}
\end{eqnarray}
\begin{eqnarray}
\rho_{e\mu-1,e\mu-1}^{\left(n\right)} & = & \bar{\Xi}^{\left(n\right)}\left(k_{\mu}^{\left(n\right)}P_{\mu}-p_{\mu-1}^{\left(n\right)}P_{\mu-1}\right)\nonumber \\
 & + & k_{f\to e}^{\left(n\right)}k_{g\to f}^{\left(n\right)}\Xi^{\left(n\right)}\bar{\Xi}^{\left(n\right)}P_{\mu-1},\label{eq:rhoeeSS}
\end{eqnarray}
where  $k_{\mu}^{\left(n\right)}$ and $p_{\mu-1}^{\left(n\right)}$
are defined in Eqs. (\ref{eq:kmu}) and (\ref{eq:pmu}). Notice that $\Theta^{(n)} \equiv \Theta^{(n)}_{\mu \mu},\bar{\Theta}^{(n)} \equiv \bar{\Theta}^{(n)}_{\mu \mu}$ and $\Xi^{(n)} \equiv \Xi^{(n)}_{\mu \mu},\bar{\Xi}^{(n)} \equiv \bar{\Xi}^{(n)}_{\mu \mu}$ do not depend on $\mu$. The remaining molecule-plasmon correlation $\rho_{f\mu,f\mu}^{\left(n\right)}$ can be calculated with
the relation $\rho_{f\mu,f\mu}^{\left(n\right)}=P_{\mu}-\rho_{g\mu,g\mu}^{\left(n\right)}-\rho_{e\mu,e\mu}^{\left(n\right)}$.

If all the molecules are identical, Eq. (\ref{eq:current})
can be reformulated as
\begin{equation}
I_{X}=N_{\text{m}}\sum_{a=g,e}\left(k_{Xa\to f}P_{a}-k_{Xf\to a}P_{f}\right),\label{eq:current-1}
\end{equation}
where $P_{a}=P_{na}$ and $P_{f}=P_{nf}$. From Eqs. (\ref{eq:rhoggSS})
and (\ref{eq:rhoeeSS}), we can easily get
\begin{align}
N_{\text{m}}P_{g} & =\bar{\Theta}\gamma_{\text{pl}}\mathcal{N}_{\text{pl}}+N_{\text{m}}k_{f\to g}k_{e\to f}\Theta\bar{\Theta},\label{eq:summation-g}
\end{align}
\begin{align}
N_{\text{m}}P_{e} & =-\bar{\Xi}\gamma_{\text{pl}}\mathcal{N}_{\text{pl}}+N_{\text{m}}k_{f\to e}k_{g\to f}\Xi\bar{\Xi}.\label{eq:summation-e}
\end{align}
Using $P_{f}=1-P_{g}-P_{e}$ and inserting Eqs. (\ref{eq:summation-g})
and (\ref{eq:summation-e}) in Eq. (\ref{eq:current-1}), we get the
Eq.(\ref{eq:current-S}) in the main text.

\section{Derivation of Rate Equations\label{sec:rate-equations-appendix}}

The equations of motion for the populations
$P_{na}\equiv\text{tr}_{\text{S}}\left\{ \hat{\rho}\left(t\right)\left|a_{n}\right\rangle \left\langle a_{n}\right|\right\} $ read
\begin{eqnarray}
\frac{\partial}{\partial t}P_{ne} & =& -k_{e\to f}^{\left(n\right)}P_{ne}+k_{f\to e}^{\left(n\right)}P_{nf}-2v_{n}\text{Im}\left\langle \left|g_{n}\right\rangle \left\langle e_{n}\right|C^{+}\right\rangle ,\label{eq:pne} \\
\frac{\partial}{\partial t}P_{ng} &=& -k_{g\to f}^{\left(n\right)}P_{ng}+k_{f\to g}^{\left(n\right)}P_{nf}+2v_{n}\text{Im}\left\langle \left|g_{n}\right\rangle \left\langle e_{n}\right|C^{+}\right\rangle .\label{eq:png}
\end{eqnarray}
The correlations $\left\langle \left|g_{n}\right\rangle \left\langle e_{n}\right|C^{+}\right\rangle \equiv\text{tr}_{\text{S}}\left\{ \hat{\rho}\left(t\right)\left|g_{n}\right\rangle \left\langle e_{n}\right|C^{+}\right\} $
satisfy the following equation
\begin{eqnarray}
\frac{\partial}{\partial t}\left\langle \left|g_{n}\right\rangle \left\langle e_{n}\right|C^{+}\right\rangle
 & = & i\left(\tilde{\omega}_{\text{pl}}^{*}-\tilde{\omega}_{n}\right)\left\langle \left|g_{n}\right\rangle \left\langle e_{n}\right|C^{+}\right\rangle \nonumber \\
 & + & iv_{n}\left(P_{ne}+\left\langle \left(\left|e_{n}\right\rangle \left\langle e_{n}\right|-\left|g_{n}\right\rangle \left\langle g_{n}\right|\right)C^{+}C\right\rangle \right)\nonumber \\
 & + & i\sum_{n'\neq n}v_{n'}\left\langle \left|e_{n'}\right\rangle \left\langle g_{n'}\right|\times\left|g_{n}\right\rangle \left\langle e_{n}\right|\right\rangle ,\label{eq:Coherence}
\end{eqnarray}
where the complex transition frequencies are defined as $\tilde{\omega}_{\text{pl}}=\omega_{\text{pl}}-i\gamma_{\text{pl}}/2$
and $\tilde{\omega}_{n}=\omega_{n}-i\left(k_{e\to f}^{\left(n\right)}+k_{g\to f}^{\left(n\right)}\right)/2$.
To obtain closed equations, we omit the short-lived correlations involving two molecules in Eq.
(\ref{eq:Coherence}). At steady state, we obtain
\begin{equation}
\left\langle \left|g_{n}\right\rangle \left\langle e_{n}\right|C^{+}\right\rangle =\frac{iv_{n}}{\tilde{\omega}_{\text{pl}}^{*}-\tilde{\omega}_{eg}}\times\left(P_{ne}+\left(P_{ne}-P_{ng}\right)\left\langle C^{+}C\right\rangle \right) \label{eq:ExpgeC}
\end{equation}
by assuming the factorization $\left\langle \left(\left|e_{n}\right\rangle \left\langle e_{n}\right|-\left|g_{n}\right\rangle \left\langle g_{n}\right|\right)C^{+}C\right\rangle =\left(P_{ne}-P_{ng}\right)\left\langle C^{+}C\right\rangle $. Inserting Eq. (\ref{eq:ExpgeC}) in Eqs. (\ref{eq:pne}) and (\ref{eq:png}),
we obtain Eqs. (\ref{eq:pneF}) and (\ref{eq:pngF}) in the main text.


\begin{thebibliography}{40}
\bibitem{Uncertainty} Heisenberg's uncertainty principle says that
the time uncertainty $\Delta t$ multiplied by the energy uncertainty
$\Delta E$ should be larger than Planck constant $\hbar$: $\Delta t\times\Delta E\geq\hbar$.
For the molecular junction, we relate $\Delta t$
with the life-time of the lead plasmon and $\Delta E$ with the uncertainty
$\Delta E_{\mathrm{pl}}$ of the plasmon excitation energy $E_{\mathrm{pl}}$.
Since the energy $E_{\mathrm{photon}}$ of emitted photons is determined
by $E{}_{\mathrm{pl}}$, the uncertainty of the photon energy $\Delta E_{\mathrm{photon}}$,
i.e. the emission line-width, is identical to $\Delta E_{\mathrm{pl}}$.
Because the plasmon damping is compensated by the excited molecules, indicated 
by the increased plasmon mean number, the plasmon life-time becomes longer. Therefore,
 the emission line-width is reversely proportional
to the plasmon mean number or the emission intensity.
\end{thebibliography}
\end{document}